\documentclass[twocolumn,english,aps,showpacs,preprintnumbers,amsmath,amssymb]{revtex4}
\usepackage{amsmath}
\usepackage[dvips]{graphicx} 
\usepackage{dcolumn}
\usepackage{bm} 
\usepackage{babel}
\newcommand{\h}{\hbar}
\newcommand{\ket}{\rangle}
\newcommand{\bra}{\langle}
\begin{document}

\title{Efficient vibrational state coupling in an optical tilted-washboard potential via multiple spatial translations and application to pulse echo}
\author{Samansa Maneshi, Jalani F. Kanem, Chao Zhuang, Matt Partlow, and Aephraim M. Steinberg}

\affiliation{Centre for Quantum Information \& Quantum Control  and  Institute for Optical Sciences,  \\Department of Physics, University of Toronto, Canada }

\date{\today}


\begin{abstract}
We measure the application of simple and compound pulses consisting of time-dependent spatial translations to coupling vibrational states of ultracold $^{85}$Rb atoms in a far-detuned 1D optical lattice. The lattice wells are so shallow as to support only two bound states, and we prepare the atoms in the ground state. The lattice is oriented vertically, leading to a tilted-washboard potential analogous to those encountered in condensed-matter systems. Experimentally, we find that a square pulse consisting of lattice displacements and a delay is more efficient than single-step and Gaussian pulses. This is described as an example of coherent control. It is striking that contrary to the intuition that soft pulses minimize loss, the Gaussian pulse is outperformed by the square pulse.
Numerical calculations are in strong agreement with our experimental results and show the superiority of the square pulse to the single-step pulse for all lattice depths and to the Gaussian pulse for lattice depths greater than $7$ lattice recoil energies.  
We also compare the effectiveness of these pulses for reviving oscillations of atoms in vibrational superposition states using the pulse-echo technique. We find that the square and Gaussian pulses result in higher echo amplitudes than the single-step pulse. These improved echo pulses allow us to probe coherence at longer times than in the past, measuring a plateau which has yet to be explained.
In addition, we show numerically that the vibrational state coupling due to such lattice manipulations is more efficient in shallow lattices than in deep lattices.  The coupling probability for an optimized single-step pulse approaches $1/e$ as the depth goes to infinity (harmonic-oscillator limit), while in shallow lattices with large anharmonicity, the coupling probability reaches a maximum value of $0.51$ for a lattice depth of $5$ recoil energies. For square and Gaussian pulses the coupling in the lattice is even stronger, reaching maxima of $0.64$ at $6$ recoil energies and $0.67$ at $5$ recoil energies, respectively. 

\end{abstract}
\maketitle 
\section{Introduction}
\label{intro}
Cold atoms in optical lattices are of great interest for the study of numerous phenomena ranging from predictions of condensed-matter theory \cite{RaizenLadder, RaizenBlockband, JJKasevich, Jaksch0, MOTGreiner} to quantum chaos \cite{RaizenQch, Brumer1, Brumer2, JalaniKR} to quantum information protocols \cite{Brennen, Jaksch, EntangleBloch}. 
The real-time adjustability of the parameters of an optical lattice make it an attractive system for exploring the controllability of quantum systems.
Control of coherent quantum states is a topic of growing excitement, and plays for instance a crucial role in the field of quantum information processing. 
In practice this control needs to be exercised in real world systems where the interaction Hamiltonian, and particularly the interactions with the environment, may not be known precisely and where the control parameters are limited in number.  We use cold atoms in a vertically oriented optical lattice as a prototype system and study how well coherence can be controlled in the resulting tilted-washboard potential, which is also relevant to a number of other physical systems, when the control parameters are restricted to the (time-dependent) phase and amplitude of this potential. 
Our particular focus is on controlling and maintaining coherence between quantized vibrational states in such a potential.

Quantized motion of atoms in optical molasses and lattices was first observed in the early 1990s 
\cite{Qmotion0, Qmotion1, Qmotion2, Qmotion3} using methods of high-resolution spectroscopy of resonance fluorescence, and stimulated Raman spectroscopy. 
Manipulations of the optical potential have been used to create coherence between vibrational states; examples are breathing-mode oscillations \cite{Raithel0} through a sudden increase of potential depth and by parametric drive, Rabi type oscillations \cite{Raizen} through periodic phase modulations of the lattice beams, and collapse and revival of wave-packet oscillations \cite{PhilipsRevival} after a sudden shift of the lattice. 
The experiments where sudden shifts of the lattice were used to couple vibrational states
\cite{PhilipsRevival, Birkl} were performed in deep lattice potentials ($U_{\circ} > 350E_{R}$), where the coupling between the lowest two vibrational states approaches the harmonic limit. The periodic phase modulations \cite{Raizen} in a very shallow lattice ($U_{\circ} \approx~6E_{R}$) did not result in large coupling either. In our experiments in moderately shallow lattices $(18 \le U_{\circ} \le 20)E_{R}$ we demonstrate how a combination of sudden shifts of the lattice and delays may result in large couplings. 
Non-adiabatic shifts of the potential have also been used in a pulse-echo experiment \cite{Birkl} to  revive wave packet oscillations. In a pulse-echo experiment, one would ideally want to completely invert 
the state populations by applying a $\pi$-pulse for a perfect revival of oscillations.  Unfortunately, the width of the bands as well as inhomogeneous broadening have prevented any one from demonstrating ideal $\pi$-pulses for all lattice quasi-momentum components simultaneously.   Echo pulses  consisting of lattice displacements and delays presented in this paper, while not perfect $\pi$-pulses, do result in partial inversion of state populations, and the strongest echo signals ever observed in such systems.   
There have also been proposals \cite{Zoller, Muga} for laser-driven vibrational state transitions by adiabatic passage along dressed energy levels for ions in harmonic and hard-wall potentials. Experimentally, Raman transitions have been successfully used to create Fock motional states for ions trapped in a harmonic potential \cite{Wineland} and for cold atoms in far-detuned 1D optical lattices \cite{Salomon}. In fact, a displacement of the lattice acts like a pair of Raman pulses though it does not result in a complete population inversion. 
Here, we present a study of the extent to which such lattice manipulations alone can result in inversion
of populations between the lowest two bands. It is interesting to note that our tilted optical lattice at depths supporting two or three bands is very similar to anharmonic three-level potentials in superconducting qubits \cite{Zagoskin, Chuang, FWilhelm, Girvin07}. In these systems, it is of fundamental interest to learn how manipulations of the potential parameters can be best used to control the quantum states of the system.   

In our experiment, first we study the efficiency of three different pulses consisting of lattice displacements and delays in coupling the atoms in the first band to the second band. Next we compare the efficiency of these pulses in reviving the oscillations of atoms that are in a superposition of vibrational states through a pulse-echo technique. Throughout this paper we refer to the first experiment as the coupling experiment and to the second experiment as the echo experiment. 

In section II we start with a description of the experiment, then in section~\ref{sec:coup} present our results in the coupling experiment and compare them with simulations. In section~\ref{sec:echo} we compare echo amplitudes after the three experimental optimized pulses. The details of our numerical calculations are provided in section III.   
\section{Experiment}
\label{experiment}
We trap $^{85}$Rb atoms in a lattice detuned $25GHz$ above the D2 transition line, $F=3 \rightarrow F^{\prime}=4$ ($\lambda=780nm$). For typical depths of $ 0.6 - 0.7\mu K$, this results in a photon scattering rate of $16Hz$, equivalent to a lifetime of $\approx~60$ms, which is much longer than the time scale of the experiment. The vertical lattice is formed by two laser beams with parallel polarizations intersecting at an angle of $\theta=49.6^{0}$, resulting in a lattice spacing of $a=\frac{\pi}{k_{L}}=0.93 \mu m$, where $k_{L}=\frac{2\pi}{\lambda}\sin(\frac{\theta}{2})$ is the lattice vector.
We load the lattice from optical molasses with the final temperature of $ \approx~10 \mu K$ and atomic density of $\approx~10^{10} {\rm atoms}/cm^3$. The molasses is kept on in the presence of the lattice (depths of $18-20 E_{R}$, where $E_{R} = \h^{2}k_{L}^{2}/2m = \h (2\pi \times 685 Hz)$ is the effective recoil 
energy for our lattice) and switched off quickly. Approximately $10\%$ of the atoms are loaded to the lattice, which corresponds to a density of $\approx~5 \times 10^5$ per lattice plane; there is no transverse confinement, and this density is low enough that we may neglect interactions between atoms.      
Due to the short coherence length of atoms in optical molasses ($\approx~60 nm$  at $ 10 \mu K$) there is no coherence between wells.  

\begin{figure}[b]
\includegraphics[scale=0.26]{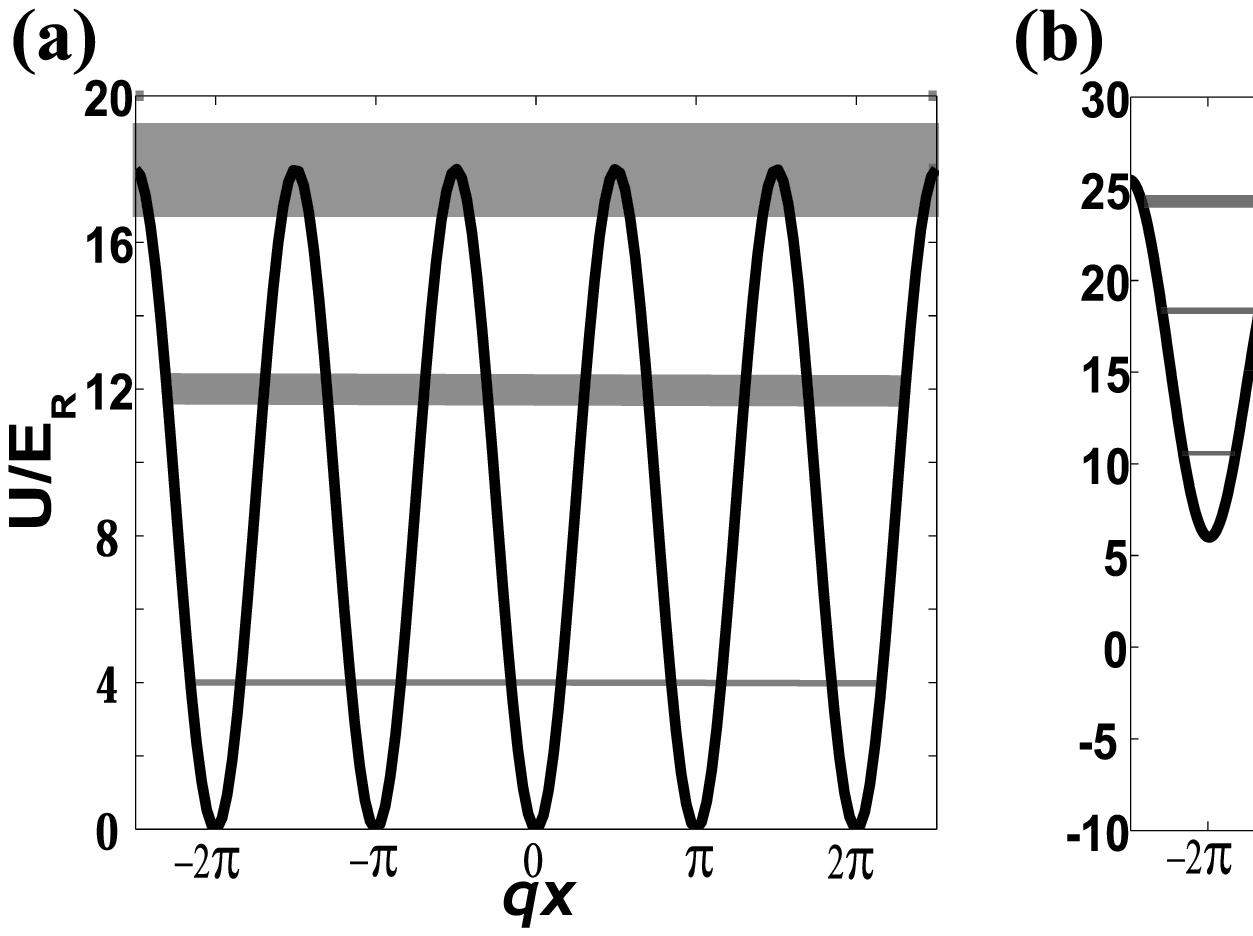}
\caption{ (a) Lattice potential $U(x) = U_{\circ}sin^{2}(k_{L}x)$ for the depth $U_{\circ} = 18E_{R}$ and the first three bands as a function of dimensionless parameter $qx=x \pi/a$, where $a$ is the lattice spacing. (b) Schematic of the tilted lattice showing $U(x)+ mgx$ and the quasi-bound states of the Wannier-Stark ladder. The energy levels are the averages of band energies and they differ from site to site by $2.86E_{R}$, where $2.86E_{R}$ is the tilt due to gravity per lattice site.}
\label{fig1}
\end{figure}

Figure~\ref{fig1}(a) is a plot of periodic potential and energy bands for a lattice depth of $U_{\circ} = 18E_{R}$ and Fig.~\ref{fig1}(b) shows the same potential in presence of gravity, where the force of gravity adds a tilt to the lattice with a magnitude of $2.86 E_{R}$ per lattice site.  
 
The atoms are prepared in the lowest vibrational band of the lattice by lowering the lattice depth adiabatically to the point that only one band is supported by the lattice. The lattice is then kept at this depth for $4 ms$ during which the unbound atoms accelerate downward and escape the lattice region due to gravity. The lattice depth is adiabatically increased back to a value that supports two bands, which corresponds in our experiment to potential heights of $U_{\circ}=18 -20 E_{R}$.  
With these operations we are able to prepare $ > 95 \%$ of the atoms in the lowest band while the remaining $ < 5\%$ are either lost from the lattice or are excited to the second band.
The Bloch oscillation period in this lattice is $\approx~530 \mu s$, but we do not observe Bloch oscillations because of the incoherent filling of the lattice, which populates all quasi-momenta with the same probability. 
Due to the tilt of lattice, the third band at lattice depths of $U_{\circ}=18-20 E_{R}$ is not bound (see Fig.~\ref{fig1}). In practice, this means that the first two states remain bound throughout the duration of our experiment, while the third state escapes via Landau-Zener tunneling in $\approx~830 \mu s$; see section \ref{model} for more details.

The depth of the lattice is calculated from the measured oscillation frequency between the lowest two bands. The technique for this frequency measurement will be described later; sample oscillations for a lattice depth of $U_{\circ}=20 E_{R}$ are shown in Fig.~\ref{fig4}. The measured average oscillation period in this figure is $T_{ave}=(187\pm2)\mu s$ and the rms width of the decaying Gaussian is $(258\pm6)\mu s$. The oscillation measurement is an ensemble measurement and the measured frequency is the average of a distribution of frequencies due to the transverse Gaussian profile of the laser beams forming the lattice. From the measured lattice beam detuning and peak intensity we calculate the potential at the center of the lattice to be $(24\pm 2)E_{R}$. This lattice depth is in the range of depths we calculate from the Gaussian rms width of the decaying oscillations. More details on the stages of preparation, operations and measurement including the adiabatic filtering and displacements of the lattice can be found in \cite{myrskog}. 
The data presented in this paper are for two lattice depths; in the coupling experiment the lattice depth is $U_{\circ}=18E_{R}$ and in the echo experiment $U_{\circ}=20 E_{R}$.  

\subsection{\label{sec:coup} Coupling Experiment}

\begin{figure}[!t]
\includegraphics[scale=0.26]{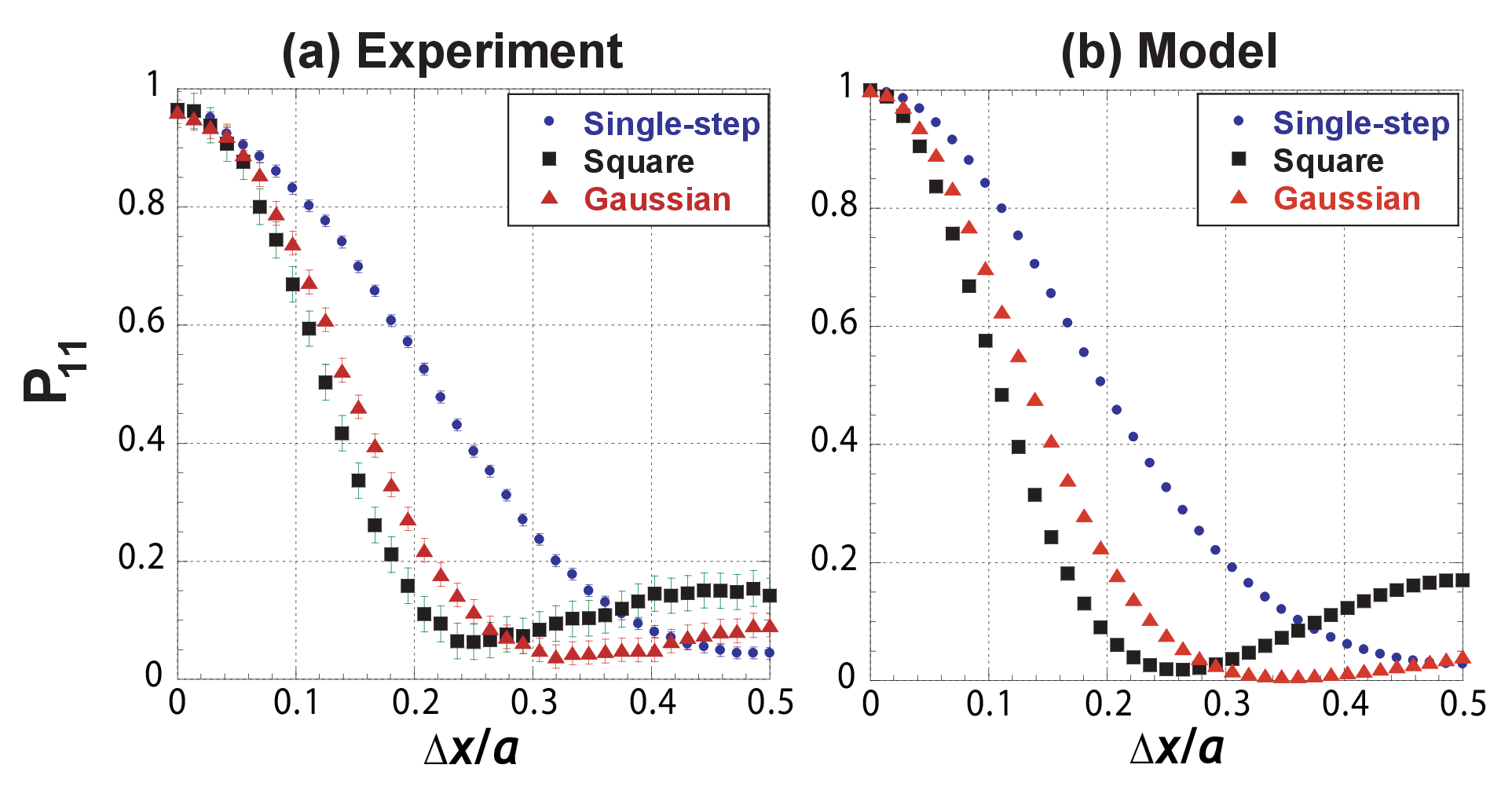}
\caption{(Color online) (a) Experimental $P_{11}$ as a function of displacement after application of single-step, square and Gaussian pulses for a lattice depth of $U_{\circ}=18E_{R}$. The temporal widths of the square and Gaussian pulses are the experimental optimized temporal widths. (b) Calculated $P_{11}$ for the three pulses, with calculated optimized temporal widths for the square and Gaussian pulses. For experimental and calculated optimized pulse parameters see table~\ref{tb1}.}
\label{fig2}
\end{figure} 
 
The three pulses used in this experiment are single-step, square and Gaussian profiles in time. 
The single-step pulse is a single lattice displacement. The square pulse consists of a displacement, delay and a displacement in the opposite direction. The Gaussian pulse consists of small displacements of the lattice in time steps of $5 \mu s$, which combined together approximate a Gaussian shaped pulse.  The lattice is displaced by changing the relative phase of the laser beams forming the lattice through acousto-optic modulators (AOMs).
The relevant parameters of each pulse are optimized by maximizing the number of atoms transferred from the first band to the second band: the single-step pulse is optimized with respect to the amplitude of single lattice displacement; the square pulse is optimized with respect to the amplitude of displacements and the delay between the displacements; and the Gaussian pulse is optimized with respect to the maximum amplitude of the lattice displacement and the FWHM temporal width of the Gaussian.
 
Starting with essentially all atoms in the lowest band ($n=1$), we measure the number remaining in band 1 and the number transferred to band 2, to infer coupling probabilities $P_{11}$ (the probability of remaining in band 1) and $P_{12}$ (the probability of being transferred to band 2). We compare these results with values calculated according to  $P_{1n}=( 1/2\pi a) \int_{-\pi/a}^{\pi/a} d q |c_{1n}|_{q}^2 $, where $ |c_{1n}|_{q}^2=|\bra \Psi_{n,q}(x)|\hat{O}| \Psi_{1,q}(x) \ket|^{2} $ is the square of the matrix element of the operator $\hat{O}$ between Bloch states of the same quasi-momentum in the first band and the states in the higher bands. 
After preparing atoms in the lowest band we apply each of the three pulses and measure $P_{11}$ and  $P_{12}$ as a function of lattice displacement. The comparisons are made for a lattice depth of $U_{\circ}=18 E_{R}$ with the average oscillation period of $\overline{T}_{12}=200 \mu s$ between the lowest two bands. Both the measurements and the calculations are done for lattice displacements in steps of $0.0139a$, equivalent to a relative phase shift of $5^{\circ}$ between the lattice beams. The pulses in the experiment and the model have been optimized independent of each other. Figure~\ref{fig2}(a) compares the experimental $P_{11}$ as a function of displacement due to the three pulses. In the case of square and Gaussian pulses the optimized experimental temporal pulse widths have been used (see table~\ref{tb1} for optimized pulse parameters), which match the temporal widths we find from theoretical optimization.
In figure~\ref{fig2}(b) we show the calculated $P_{11}$ as a function of displacement after the three pulses. There is a very good agreement between the experiment and the model for $P_{11}$.  

\begin{figure}[t]
\includegraphics[scale=0.26]{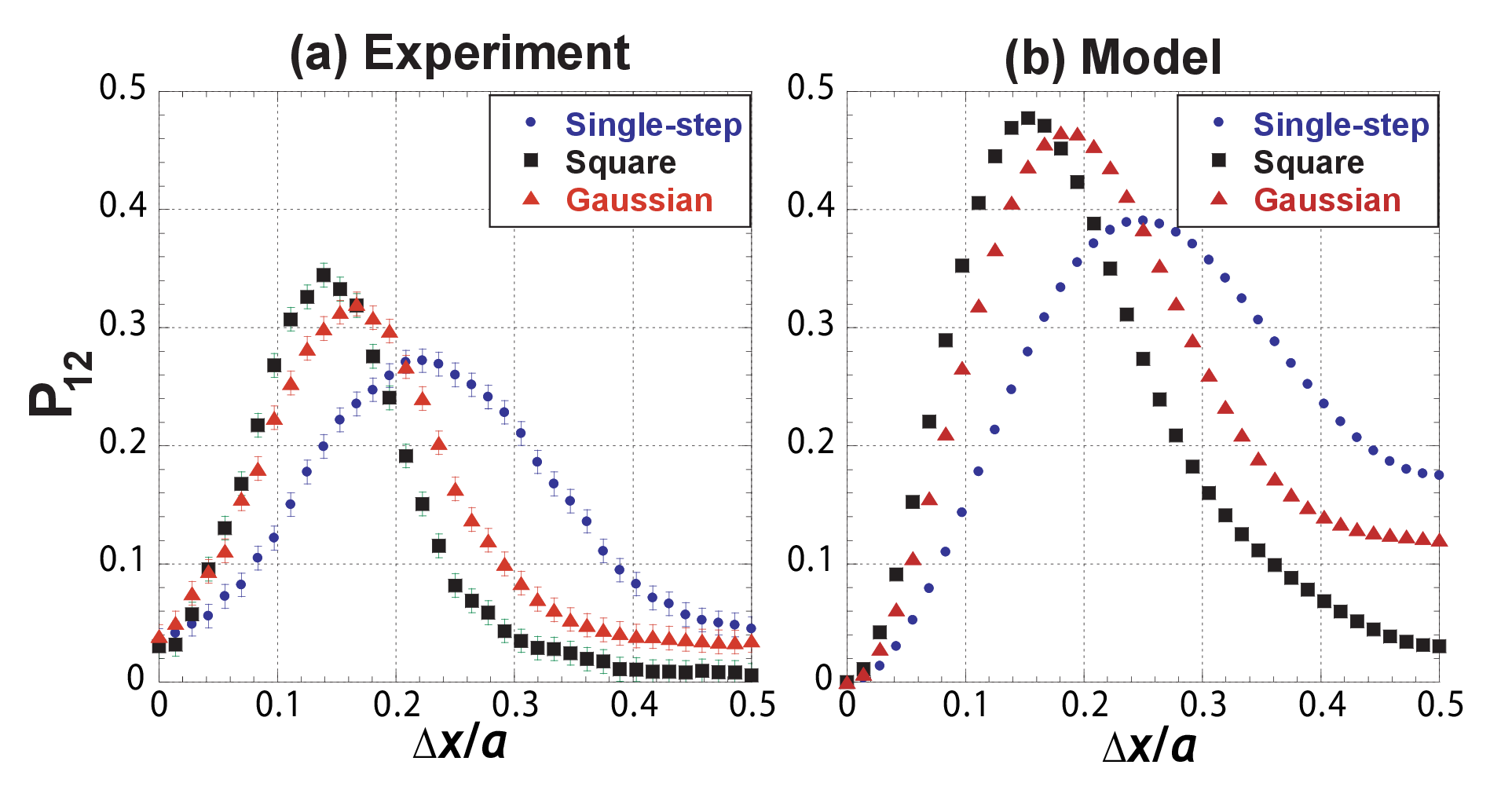}
\caption{(Color online) (a) Experimental $P_{12}$ as a function of displacement after application of single-step, square and Gaussian pulses for a lattice depth of $U_{\circ}=18E_{R}$. (b) Calculated $P_{12}$ for the three pulses. Experimental and calculated temporal pulse widths are the same as in Fig.~\ref{fig2}.}
\label{fig3}
\end{figure} 

\begin{table*}[t]
\begin{ruledtabular}
  \begin{tabular}{cccc|ccc}
  &\multicolumn{3}{c}{\it{Experiment}}  &\multicolumn{3}{c}{\it{Model}}\\
  		$Optimized$  $Pulse$      &  $A_{pulse}$   &  $W_{pulse}$ & $P_{12} $  &  $A_{pulse}$   &  $W_{pulse}$  & $P_{12}$ \\
  		\hline
			Single-step &  $(0.22\pm 0.04)a$ &   -  &  $0.26\pm 0.01$  &    $0.250a$   &   -     & $0.391$   \\
			Square &  $(0.14\pm 0.02)a$ &  $(0.35 \pm 0.03) \overline{T}_{12}$   &  $0.33\pm 0.01$ &  $0.154a$ & $0.350 \overline{T}_{12}$  & $0.477$  \\   
			Gaussian &  $(0.16\pm 0.02)a$ & $(0.29 \pm 0.03)\overline{T}_{12}$ &   $0.31\pm 0.01$ & $0.186a$  &  $0.294\overline{T}_{12}$  &  $0.466$  
		\end{tabular} 
\end{ruledtabular}
\caption{ Optimized pulse parameters (spatial amplitude $A_{pulse}$ and temporal width $W_{pulse}$) and population transfer to the second band $P_{12}$ after each pulse for a lattice depth of $U_{\circ}=18E_{R}$, with all atoms initially in the lowest band. The temporal width reported for the Gaussian pulse is the FWHM width.  
$A_{pulse}$ is given in terms of the lattice spacing $a=0.93 \mu m$, and $W_{pulse}$ in terms of average oscillation period between the lowest two bands $\overline{T}_{12}=200 \mu s$. The error bars for $A_{pulse}$ and $W_{pulse}$ indicate the uncertainties in our determination of the optimal pulse parameters, while those for $P_{12}$ are the direct measurement errors.}
\label{tb1}
\end{table*}

Figures \ref{fig3}(a) and \ref{fig3}(b) show the experimental and calculated $P_{12}$, the coupling to the second band after each pulse as a function of displacement. The temporal pulse widths are the same as in Fig. \ref{fig2}.
In the case of the single-step pulse, we find the largest transfer to the second band when the lattice displacement is $ 0.22 a=205 nm $, where $a=0.93 \mu m$ is the lattice spacing. For the square pulse, the optimum occurs for a delay of $0.35\overline{T}_{12}$, and a displacement of $0.14a = 130nm$.
We find the optimum parameters for the Gaussian pulse to be a maximum displacement of $0.16a = 149nm$ and a FWHM temporal width of $0.294\overline{T}_{12}$. Table~\ref{tb1} is a summary and comparison of experimental results with our numerical model, which is described in detail in section \ref{model}. The temporal widths of the experimental pulses match the model exactly, while the amplitudes of the optimal pulses in the experiment are smaller than the ones calculated in the model. We believe this is mainly due to the lattice inhomogeneity. Despite this difference, the optimized pulse parameters are remarkably similar for the experiment and the model. The main difference is in the maximum couplings to the second band, which in the case of experiment are $\approx~70\%$ of the calculated ones for the three pulses (see Fig.~\ref{fig3} and Table~\ref{tb1}). The loss from the lattice after each pulse due to couplings to higher bands not bound in the latice is larger in the case of experiment than in the calculation, which we attribute to experimental imperfections. This difference becomes more pronounced as the magnitude of lattice displacement increases to the maximum of $\Delta x/a =0.5$ (see Fig.~\ref{fig3}).   

\begin{figure}[!b]
\includegraphics[scale=0.26]{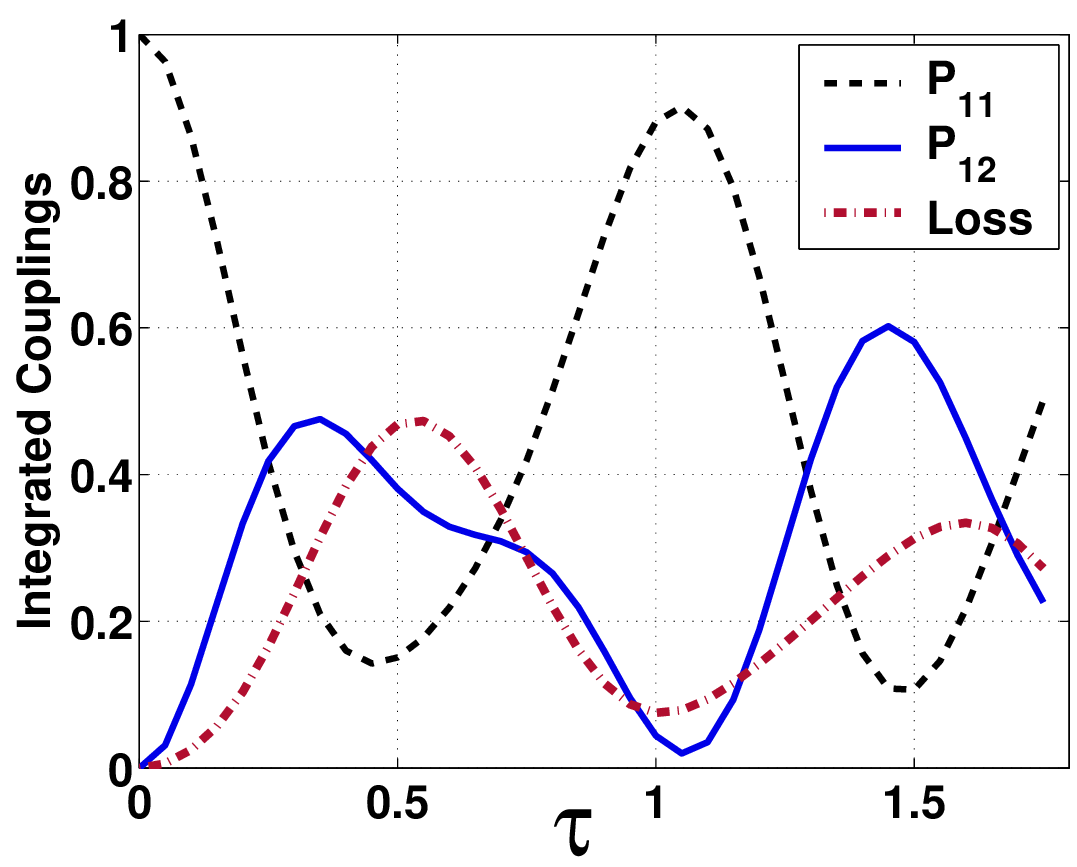}
\caption{(Color online) Calculated couplings $P_{12}$, $P_{11}$ and loss vs. scaled delay for the optimized square pulse at fixed $\Delta x =0.16a$ in a lattice depth of $U_{\circ}=18E_{R}$. The ground and excited states populations coherently oscillate as a function of delay.}
\label{fignew4}
\end{figure}

Overall, both the experimental results and the calculations show the same trend in the performance of the pulses, i.e., both show that the square pulse results in higher coupling to the second band than the single-step and Gaussian pulses. The square pulse is an example of coherent control. 
The main feature of a coherent control scheme is the ability to steer a process toward a desired outcome. In the simplest case, two different processes can lead to the same final state and interference between these processes allows one to manipulate the branching ratio into this final state by controlling the relative phase of the two processes \cite{BrumerShapiro}. 
For the pulses discussed in this paper, both the square and Gaussian pulses introduce a phase factor in the superposition state through time delays. Consider the simpler of the two pulses, the square pulse,  in which the time delay is introduced in one step. Each displacement in the square pulse couples states $1$ and $2$, analogous to a beam splitter in a Mach-Zehnder interferometer. The appropriate choice of the relative phase between the two displacements maximizes the transfer to state $2$. Figure~\ref{fignew4} shows the integrated couplings $P_{12}$, $P_{11}$ and loss vs. scaled delay after the square pulse at the fixed displacement of $\Delta x =0.16a$. The plot clearly shows the coherent oscillations of the populations as a function of delay, demonstrating that the final populations are determined by interference between the two paths. 

\begin{figure}[!b]
\includegraphics[scale=0.26]{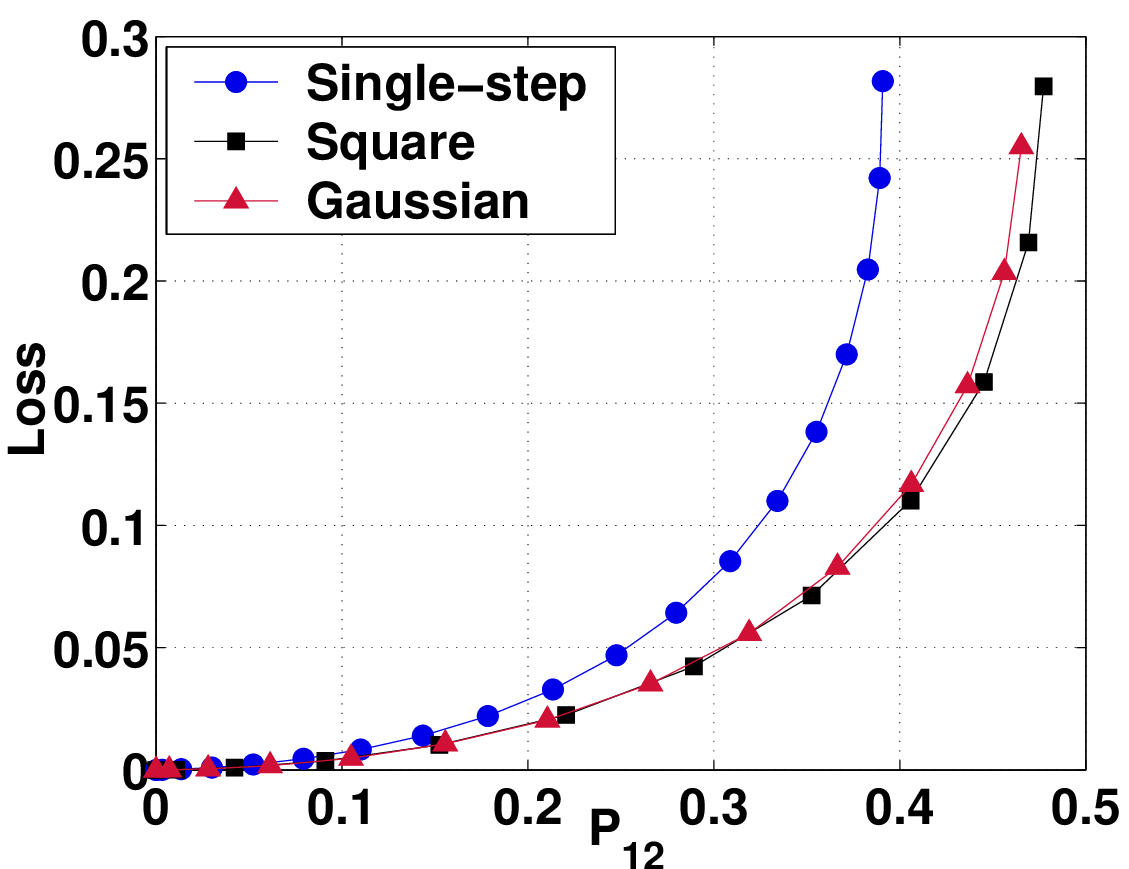}
\caption{(Color online) Calculations: Loss vs. coupling $P_{12}$ after the square, Gaussian and single-step pulses in a lattice depth of  $U_{\circ}=18E_{R}$. For the same coupling $P_{12}$, the square pulse results in less loss than the single-step and Gaussian pulses.}
\label{fignew5}
\end{figure}

Not only can phase be used to enhance transfer into the desired state via constructive interference, but it can simultaneously minimize loss through destructive interference into the undesired state. As 
Fig.~\ref{fignew4} shows, the coupling $P_{12}$ is peaked at $\tau < 0.5$, while the coupling to higher levels (which result in loss) peaks later. This enables us to choose an optimal time delay for high coupling $P_{12}$ but low loss. This result is analogous to techniques which have been used in other  experiments to optimize coupling while minimizing loss \cite{CornellPrentiss}.  

To elucidate the utility of such coherent control for optimizing the branching ratio, 
we plot in Fig.~\ref{fignew5} loss versus coupling $P_{12}$ for the single-step, square and Gaussian pulses. The displacement magnitude increases as the curves are traced counterclockwise from zero.
For the same coupling $P_{12}$, the square pulse results in less loss than the single-step and Gaussian pulses. As $P_{12}$ grows to $0.39$ (the maximum achievable with a single-step pulse in a lattice depth of $18E_{R}$), the loss due to the single-step pulse becomes nearly $3$ times larger than that due to a square pulse; for larger $P_{12}$, the single-step pulse is insufficient, regardless of loss. 
The square and Gaussian pulses have similar effect but still the square pulse outperforms the Gaussian pulse. The surprising result here is that the square pulse involving a pair of hard pulses does a better job of minimizing loss for a given transfer probability into the excited state than the Gaussian soft pulse despite the frequency selectivity of the Gaussian pulse \cite{Chuang}. 

\subsection{\label{sec:echo}Pulse-echo Experiment}
To measure the efficiency of each pulse in reviving oscillations, we first prepare atoms in the lowest band, then displace the lattice by $a/6 = 155nm $, where $a$ is the lattice spacing. The displacement of the lattice puts atoms in a superposition of the lowest two bands. This displacement magnitude is chosen as a trade-off between optimum coupling and loss of atoms from the lattice due to a single shift. (We have verified experimentally that an initial displacement of this size results in larger original oscillation amplitude and a larger echo amplitude than any other initial displacement.)  
We then map the phase evolution of the superposition state into the population evolution of the lowest band $|1\ket$ by another displacement of the lattice. Mathematically, these operations can be written as
\begin{equation}
P_{f}(t)=\bra 1|D^{\dagger}(|\alpha|)R(\omega t)\rho R^{\dagger}(\omega t)D(|\alpha|)|1\ket,
\label{oscilln}
\end{equation}
where $\rho$ is the initial density matrix describing the superposition state,  
$D(|\alpha|)=exp(-i \hat{P}|\alpha|/\hbar)$ is the displacement operator, and $R(\omega t)=exp(-i Ht/\hbar)$ is the time evolution operator.    
This technique is analogous to the Ramsey technique of separated oscillatory fields \cite{Ramsey}. The number of atoms left in the lowest band after the above operations is measured by lowering the lattice until all other bands become untrapped, and counting the remaining atoms using fluorescence imaging \cite{myrskog}.  

\begin{figure}[t]
\includegraphics[scale=0.30]{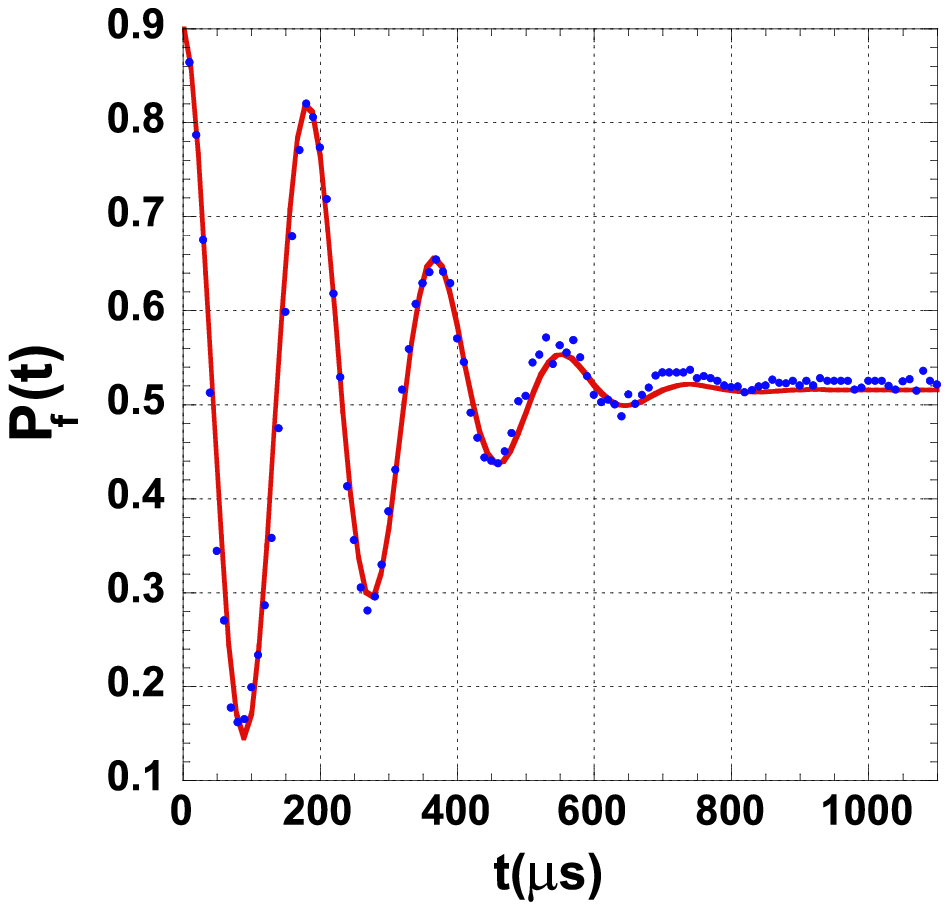}
\caption{(Color online) Experimental data: Evolution of $P_{f}(t)$ as a function of time. The curve is a fit to data (sinusoid with Gaussian decay). $T_{ave}=(187\pm2)\mu s$ and rms width of Gaussian is $(258\pm6)\mu s$}
\label{fig4}
\end{figure} 
  
\begin{table*}[t]
\begin{ruledtabular}
\begin{tabular}{ccccc|c}
	& & & \multicolumn{2}{c}{Experiment}  &\multicolumn{1}{c}{Model}\\
 Pulse  & $A_{pulse}$  & $W_{pulse}$ &  Echo amplitude & $P_{n \ge 3}$ &  $P_{n \ge 3}$\\
\hline
 Single-step 	& $(0.29 \pm 0.05)a$  & -  & $0.19 \pm 0.01$   & $71\%$ &  $52\%$ \\
 Square & $(0.17 \pm 0.03)a$ & $(0.40 \pm 0.05)\overline{T}_{12} $  & $0.23 \pm 0.01$  & $70\%$ &  $48\%$ \\
 Gaussian & $(0.17 \pm 0.03)a $ &  $(0.31\pm 0.06)\overline{T}_{12}$  & $0.24 \pm 0.01$  & $55\%$ &   $39\%$
 \end{tabular}
\end{ruledtabular}
\caption{Echo amplitude and number of atoms lost due to each pulse centered at $t_{\circ}=1040 \mu s$ for a lattice depth of $U_{\circ}=20E_{R}$ and the initial 'coherent' state prepared by the $a/6$ displacement of the lattice. The echo amplitude is the ratio of the amplitude of the echo Gaussian envelope to $0.466$,  the amplitude of the original oscillations. $A_{pulse}$ is the maximum lattice displacement, $W_{pulse}$ is the temporal width of the pulse with $\overline{T}_{12} = 188 \mu s$ at $20E_{R}$ lattice depth, and $P_{n \ge 3}$ is the number of atoms in higher bands. The error bars for $A_{pulse}$ and $W_{pulse}$ indicate the uncertainties in our determination of the optimal pulse parameters, while those for the echo amplitudes are the direct measurement errors. The last column is the calculated $P_{n \ge 3}$ for the experimental pulse parameters.}
\label{tb2}
\end{table*}

Figure~\ref{fig4} shows a typical measurement of the evolution of $P_{f}(t)$. The oscillations have an average period of $T_{ave}=(187\pm2)\mu s$ with rms width of $(258\pm6)\mu s$.
We believe much of the dephasing of oscillations is due to the transverse Gaussian profile of the laser beams, which results in an inhomogeneous distribution of lattice depths and therefore an inhomogeneous distribution of oscillation frequencies. The oscillations can be restored by inverting the populations of the two states, i.e., ideally by applying a $\pi$-pulse. This results in a revival of oscillations known as the echo signal. Spin-echo \cite{Hahn} in nuclear magnetic resonance and photon-echo \cite{Eberly} in optical resonances are famous examples of this type of revival.

\begin{figure}[b]
\includegraphics[scale=0.26]{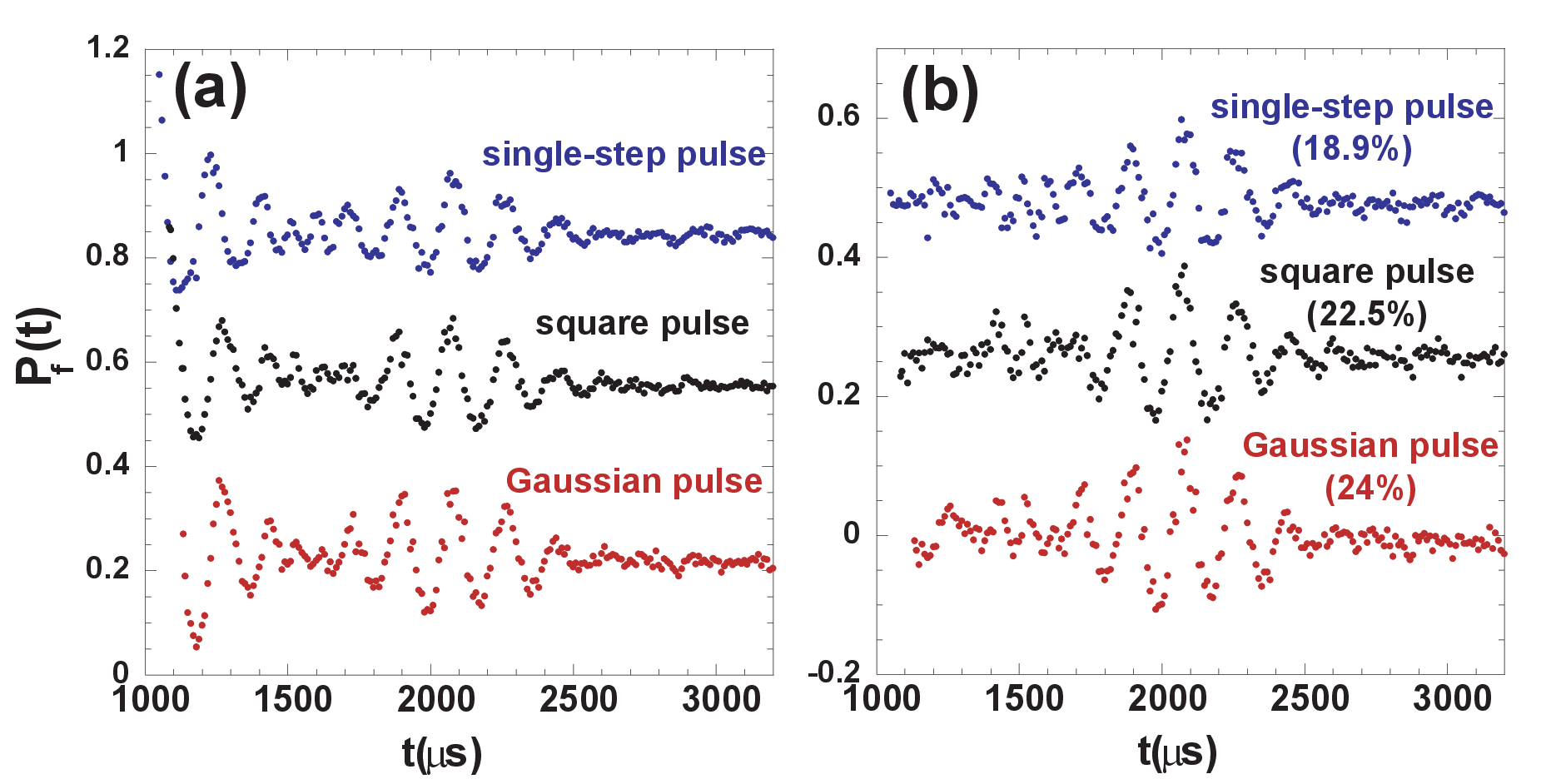}
\caption{(Color online) (a) Full echo curves after applying each pulse centered at time $t_{0}=1040 \mu s$. All echo signals are centered at $2t_{0}=2080 \mu s$. The initial oscillations are due to the echo pulse itself. (b) Subtracted echo curves where the initial oscillations due to pulses measured after $4ms$ have been subtracted from the curves in (a). All curves have been displaced vertically for clarity.}
\label{fig5}
\end{figure}

The echo pulse is applied when there are no observable oscillations, usually after 3-4 periods of oscillations. For an echo pulse centered at time $t_{\circ}$, a revival of oscillations or the echo signal centered at time $2t_{\circ}$ is observed. Figure~\ref{fig5}(a) shows the full trace of the lowest band population evolution after applying different pulses. All pulses are centered at $t_{\circ}=1040 \mu s$. The initial oscillations on each curve are due to the echo pulse itself since it is not a perfect $\pi$-pulse and creates its own coherence. These oscillations dephase on the same time scale as the original oscillations due to the same inhomogeneous broadening. 

Due to the existence of original coherence, there is a re-phasing of oscillations and we observe an echo of the original oscillations centered at $2t_{\circ}=2080 \mu s$. The echo amplitude is extracted by subtracting the oscillations caused by the echo pulse itself. These oscillations are measured with each pulse applied at a time when there is no coherence left in the system and therefore we do not expect any revival of oscillations. In this experiment, each pulse is applied at $t_{\circ}=4 ms$ and the measured oscillations are then subtracted from the curves in Fig.~\ref{fig5}(a). Figure~\ref{fig5}(b) shows the resultant echo curve after each pulse. 

The echo amplitude in the case of a single-step pulse at the earliest time we can measure, $t=2080\mu s$, is $\approx~0.19$ times the original oscillation amplitude, roughly an order of magnitude larger than the one observed in \cite{Birkl}. The main reason for this difference is that our lattice is an essentially conservative potential and the photon scattering rate is negligible during the experimental run time ($ \Gamma_{sc} = 16Hz$). In \cite{Birkl}, the lattice is dissipative (cooling in the lattice) and atoms scatter a large number of photons ($\Gamma_{sc} =2.5 MHz$). The lower echo amplitude in \cite{Birkl} is mainly due to photon scattering. By the time we measure the echo amplitude, i.e., at $t=2080 \mu s  \sim 11 $ oscillation periods, the atoms have scattered $0.03 $ photons while in \cite{Birkl}, by the time the echo amplitude is measured, i.e., at $t=64 \mu s \sim 12$ oscillation periods, the atoms have scattered $160$ photons. 

Some other principal differences in the two experiments are as follows. Our experiment is performed in relatively shallow lattices with lattice depths ranging $18-20 E_{R}$, while the lattice depths in \cite{Birkl} are in the range of $350-830 E_{R}$, where the coupling probability approaches the harmonic potential limit. 
We prepare atoms in the lowest band and measure the oscillation period between the lowest two bands while in \cite{Birkl} many bands are initially thermally populated. We measure the coherence between two energy levels and anharmonicity does not play a role in the decay of oscillations. In the Hannover experiment \cite{Birkl} many different anharmonic energy levels are coupled through lattice displacements and the observed signal is a measure of mean atom position as the coherences between all of these levels evolve. 
In addition, the two experiments measure different things. In \cite{Birkl} the authors measure the expectation value of the center of mass motion of the atoms, $\bra x \ket$. We measure all state populations and can use this to perform complete reconstruction of the quantum state \cite{myrskog}).
 
\begin{table*}[!t]
\begin{ruledtabular}
\begin{tabular}{ccccc|cccc}
    &	 \multicolumn{4}{c}{Optimized Echo pulse}  &  \multicolumn{4}{c}{ Optimized Coupling pulse} \\
 Pulse &$A_{pulse}$ \& $W_{pulse}$  &  $P_{11}$  &  $P_{22}$ &  $P_{12}$ & $A_{pulse}$ \& $W_{pulse}$ &  $P_{11}$ &  $P_{22}$ &  $P_{12}$\\
\hline
 Single-step  & $[ 0.29a,   -]$  & $0.201$  & $0.052$   & $0.356$ &  $[  0.24a,    -]$ & $0.334$ & $0.018$ & $0.388$ \\ 
 Square & $[0.167a, 0.40\overline{T}_{12}]$ & $0.116$  & $0.095$  & $0.415$ &  $[0.160a, 0.32\overline{T}_{12}]$  & $0.234$ & $0.005$ & $0.470$ \\
 Gaussian & [$0.167a, 0.31\overline{T}_{12}]$  & $0.298$  & $0.020$ & $0.451$ & $[0.197a, 0.251\overline{T}_{12}]$ & $0.250$ & $0.013$ & $0.462$
\end{tabular}
\end{ruledtabular}
\caption{Calculated coupling probabilities after single-step, square and Gaussian pulses for the pulse parameters used in the echo experiment (optimized echo pulse) and for the pulse parameters that give maximum $P_{12}$ (optimized coupling pulse) in a lattice depth of $20E_{R}$. $P_{11}$ and $P_{22}$ are calculated considering all the population in the lowest or the next higher band. $P_{12}=P_{21}$ is the coupling probability between the two bands.}
\label{tb3}
\end{table*}

Table~\ref{tb2} gives a summary of pulse parameters, echo amplitudes and number of atoms lost from the lattice due to optimized pulses. In this case, each pulse is optimized with respect to the amplitude of the echo. The envelope of the echo signal is fit to a Gaussian and the ratio to the original oscillation amplitude of $0.466$ is presented in table~\ref{tb2}.
The square and Gaussian pulses result in larger echo amplitudes than the single-step pulse. The larger absolute echo amplitude after the Gaussian pulse is due to lower loss of atoms from the lattice than when the other pulse shapes are used.  A normalization of the echo amplitudes with respect to the number of atoms left in the lattice (sum of the population of the two bands bound in the lattice) shows that the square pulse gives the largest relative echo amplitude. More atoms are lost from the lattice when we apply the echo pulse than when we prepare atoms in the lowest band and apply the optimized pulses. This is simply because at the time the echo pulse is applied there is a good fraction of atoms in the second band and the coupling from this band to higher bands is larger than the coupling from the lowest band to higher bands. For comparison to calculation, the last column in Table~\ref{tb2} shows the calculated values for the coupling to higher bands. The calculated values show a similar trend to the experimental values in loss of atoms due to pulses. 

The amplitude and temporal widths of the experimentally optimized echo pulses are listed in Table~\ref{tb3} and compared to the pulse parameters we calculate for optimum coupling in a lattice depth of $20E_{R}$.  
The experimental echo pulse amplitudes are in agreement with calculation within the experimental uncertainties, but the agreement is not as good as in the case of the coupling experiment we discussed in section~\ref{sec:coup} (see Table~\ref{tb1}). The largest discrepancy is in the temporal width of the square pulse. The main difference between the two experiments, the coupling and echo experiments is in the initial conditions. In the pulse-echo experiment, at the time the pulse is applied there are atoms occupying both lowest bands whereas in the case of coupling experiment 
initially all atoms are in the first band. An ideal echo pulse is the one that inverts populations of the lowest two bands. The optimum pulses discussed in section~\ref{sec:coup} result in maximum coupling to the second band, but they also leave a good fraction of atoms in the first band.  

\begin{figure}[b]
	\includegraphics[scale=0.30]{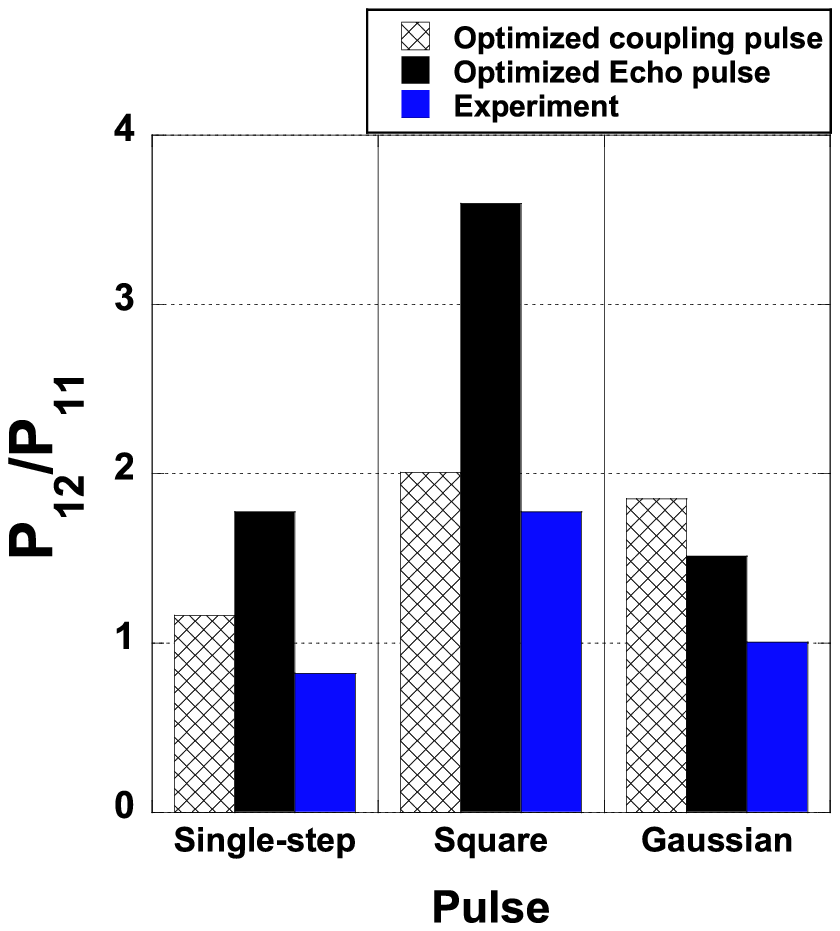}
\caption{(Color online) Bar plot of calculated ratio $P_{12}/P_{11}$ for the pulse parameters that result in maximum $P_{12}$ (optimized coupling pulse, checkered pattern bars) and the pulse parameters that result in maximum echo amplitude (optimized echo pulse, solid black bars). The blue bars show the experimental results.}
\label{fig6}
\end{figure}

Table~\ref{tb3} lists the calculated number of atoms in each band after the single-step, square and Gaussian pulses for experimental optimized pulse parameters used in the echo experiment and compares those to the number of atoms in each band if the pulse parameters for optimum coupling $P_{12}$ are used. $P_{11}$ and $P_{22}$ are calculated coupling probabilities considering all atoms in band $n=1$ and band $n=2$ respectively. $P_{12}=P_{21}$ is the coupling probability between the two bands. 
In the case of single-step and square pulses, with some compromise on the coupling, the pulse parameters used in the echo experiment leave fewer atoms in the lowest band and achieve a better population inversion and a larger echo amplitude. Figure~\ref{fig6} is a plot of the ratio $P_{12}/P_{11}$ to illustrate this fact. As can be seen in the figure, this argument does not hold for the Gaussian pulse.  We have no intuitive explanation for this interesting observation.
Figure~\ref{fig6} also predicts when starting with atoms in the lowest band, after
the experimentally optimized echo square pulse $78\%$ of the remaining atoms will be in the second band ($P_{12}/(P_{11}+P_{12})=0.78 $). Experimentally, we find this number to be $\approx~0.63$ after the optimized echo square pulse.   

\begin{figure}[b]
			\includegraphics[scale=0.30]{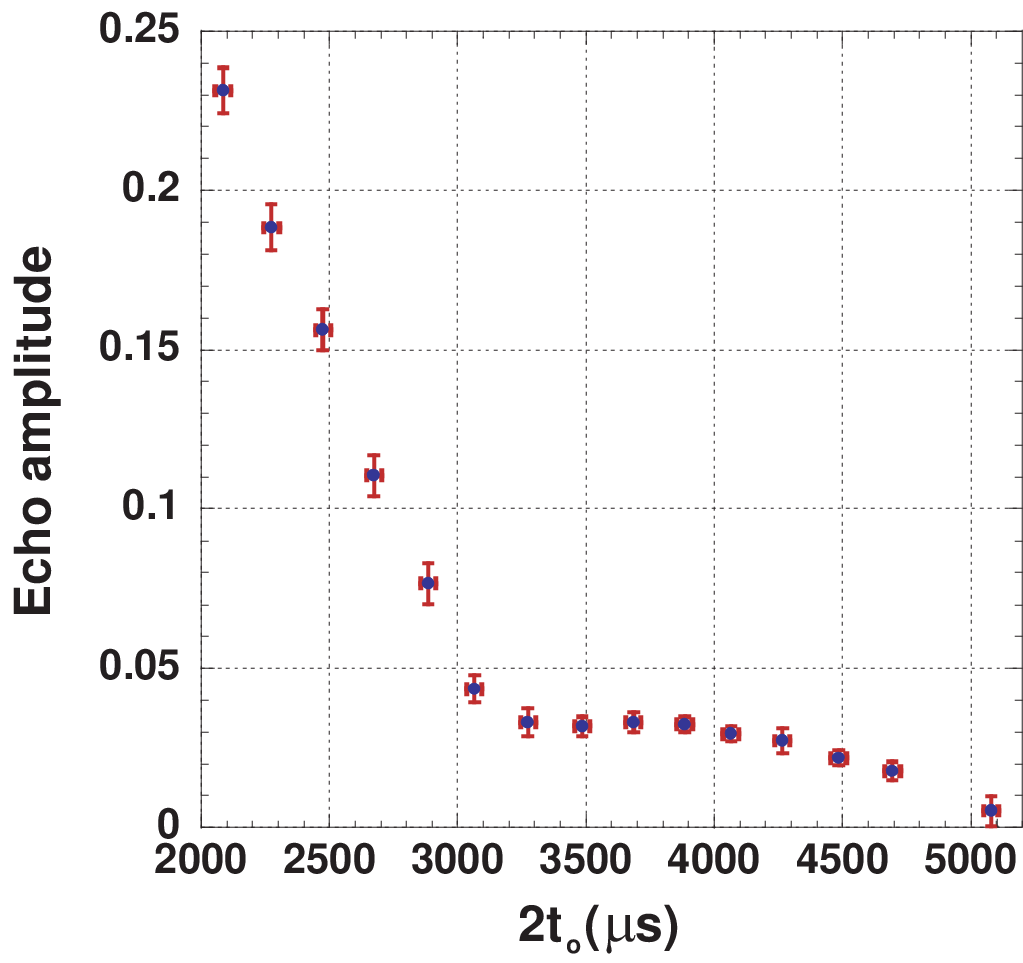}
\caption{(Color online) Echo amplitude as a function of time delay. The pulse used for the echo is the optimized experimental square pulse for the lattice depth of $U_{\circ}=20 E_{R}$, and is centered at time $t_{\circ}$; the echo is observed at $2t_{\circ}$.}
\label{fig7}
\end{figure}

We also measure how echo amplitude changes as a function of time. Figure~\ref{fig7} shows the measured echo amplitudes centered at $2t_{0}$ after applying the optimized square pulse (for pulse parameters see Table~\ref{tb2}) at time $t_{0}$. Initially, the echo amplitude decays with an exponential decay constant of $ (720 \pm 70)\mu s$, then it stays approximately constant over $1ms$ followed by a final decay. Further experiments are planned to understand this behavior, which we believe provides information about the temporal correlation properties of the instantaneous well depth experienced by the atoms as they drift through a spatially inhomogeneous lattice. 
In the 1D lattice, atoms are free to move in the transverse plane with an average velocity, $v_{rms} \propto \sqrt{T}$, where $T$ is the temperature of atoms after molasses cooling. For example, a temperature of $\approx~10 \mu K$, a typical temperature after molasses cooling, corresponds to $v_{rms} \approx~3.16 cm/s$. An order of magnitude calculation for the time it takes for an atom with this transverse velocity to accumulate a $\pi$-phase shift across a distance of $1mm$, the rms radius of the lattice beams gives a decay time of $\approx~1.7 ms$. 

\section{Modeling}
\label{model}

\begin{figure}[t]
			\includegraphics[scale=0.25]{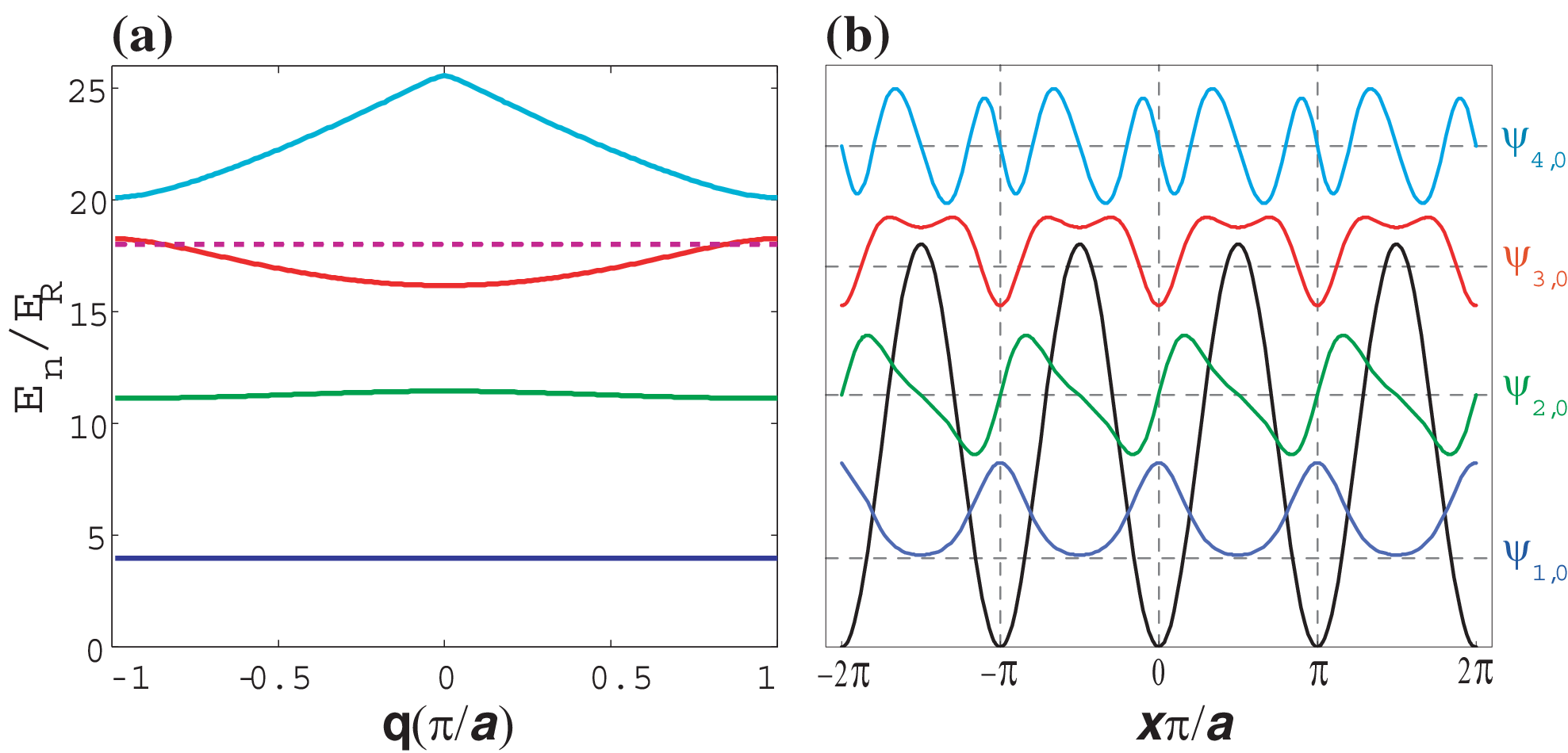}
\caption{(Color online) (a) Band structure of the periodic potential with $U_{\circ}=18E_{R}$, showing the first four bands. The dashed line indicates the depth of lattice.(b) $\psi_{n,q=0}(x\pi/a)$, the Bloch functions for quasi-momentum $q=0$ in the first four bands as a function of position. Lattice spacing is $\pi$ with the central well extending from $-\pi/2$ to $\pi/2$. The solid black curve is the lattice potential; the vertical dotted lines specify the center of wells and the horizontal dotted lines are drown for the illustration of wavefunctions. The Bloch functions have been rescaled and displaced vertically for clarity.}
\label{fig8}
\end{figure}

In modeling our experiment, we start by calculating the eigenstates of the Hamiltonian $H=P^{2}/2m + U(x)$, where $U(x)=U_{\circ} cos^{2}(k_{L}x)$ is the periodic potential. $U_{\circ}=sE_{R}$ is the lattice depth, where $E_{R}=\frac{\hbar^2k_{L}^2}{2m}$ is the recoil energy, and $k_{L}$ is the lattice vector. The eigenstates of the periodic potential are Bloch functions, $\Psi_{n,q}=e^{iqx}u_{n,q}(x)$, where $q$ is the quasi-momentum and $n$ is the band index \cite{AnM}.
Figure~\ref{fig8}(a) is the band structure for a lattice depth of $18 E_{R}$, a typical depth in our experiment. As an example of eigenstates, in Fig.~\ref{fig8}(b) we plot the $q=0$ Bloch functions of the first four bands.

Since in the experiment we prepare atoms in the lowest band of the lattice from a thermal cloud, all quasi-momenta of the lowest band are populated incoherently and with the same probability. 
This mixed state can be decomposed as an incoherent mixture of Wannier functions localized in each well. In one dimension, the Wannier function centered at $x=0$ is defined as \cite{Korsch}    

\begin{equation}
w_{n,l}(x)=\int_{-\pi/a}^{\pi/a} \mathrm{d}q \psi_{n,q}(x)e^{-iqla}
\end{equation}

Wannier functions are exponentially localized to each lattice site and they satisfy the relation $w_{n,l}(x)=w_{n,l}(x-a)$. In a tilted lattice, the eigenstates of the Hamiltonian 
$H=P^{2}/2m + V(x) + F x$ are the metastable Wannier-Stark states, which form the discrete Wannier-Stark ladder \cite{Korsch}. The Wannier-Stark states for the lowest lying states are localized to each well for small values of $F$. The lifetimes of the Wannier-Stark states are determined from the band-dependent Landau-Zener transition rates \cite{Zener}, $\Gamma_{LZ} = \nu_{B} exp(-\pi^{2} a E_{G}^{2}/h^{2} g n)$, where $\nu_{B} = Fa/h$ is the Bloch frequency, $a$ is the lattice spacing, $E_{G}$ is the band-gap, $g$ is the acceleration due to gravity, and $n$ is the band index for a transition from band $n$ to band $n+1$. For the lowest two states in our lattice depth of $18E_{R}$ and the tilt per lattice site of $2.86E_{R}$, the Landau-Zener transition rate for the $n=1 \to n=2$ transition is $\approx~4 \times 10^{-7} Hz$ and for $n=2 \to n=3$ it is $6 Hz$. This results in lifetimes of $2.5 \times 10^{6} sec$ and $166 ms$ respectively for these transitions, which are larger than the time scale of our experiment. For the $n=3 \to n=4$ transition the rate is $1.2 \times 10^{3} Hz$ which corresponds to a lifetime of $830 \mu s$. Since both the Bloch oscillation period ($530 \mu s$), and the Landau-Zener lifetimes are much longer than the temporal widths of the pulses used here, we believe that the tilt of the lattice may be neglected. This is borne out by the agreement of our model with experiment, and has more recently been confirmed by comparison with more complete calculations including gravity (to be published elsewhere).
 
A periodic Hamiltonian $H=P^{2}/2m + U_{\circ} cos^{2}(k_{L}x)$ conserves quasi-momentum. In addition, an abrupt displacement of the lattice does not change quasi-momentum. We therefore operate on each Bloch function in the first band with a pulse $\hat{O}$, then calculate the square of the overlap of the transformed Bloch function with the Bloch function $\Psi_{n,q}(x)$ of the same quasi-momentum in higher bands. Since we model the initial state as an incoherent sum over all quasi-momenta, we calculate the average transfer probability over all values of quasi-momentum, 

\begin{subequations}
	\begin{equation}
	P_{1n} =\frac{1}{2\pi a} \int_{-\pi/a}^{\pi/a} d q |c_{1n}|_{q}^2, 
	\end{equation}
where 
	\begin{equation}
	|c_{1n}|_{q}^2  \equiv |\bra \Psi_{n,q}(x)|\hat{O}| \Psi_{1,q}(x) \ket|^{2}
	\end{equation}
\label{P1nNc1n}
\end{subequations}
is the square of the matrix element of the operator $\hat{O}$ for each quasi-momentum.     
For example, to find the coupling due to a single-step pulse consisting of a lattice displacement, we calculate $|c_{1n}|_{q}^2=|\bra \Psi_{n,q}(x)|\Psi_{1,q}(x-\Delta x) \ket|^{2}$ for each quasi-momentum and integrate over all quasi-momenta to find the total coupling probability. In calculations, seven bands have been included and we have checked that including more bands (up to $20$) does not change the result.

\begin{figure}[t]
			\includegraphics[scale=0.25]{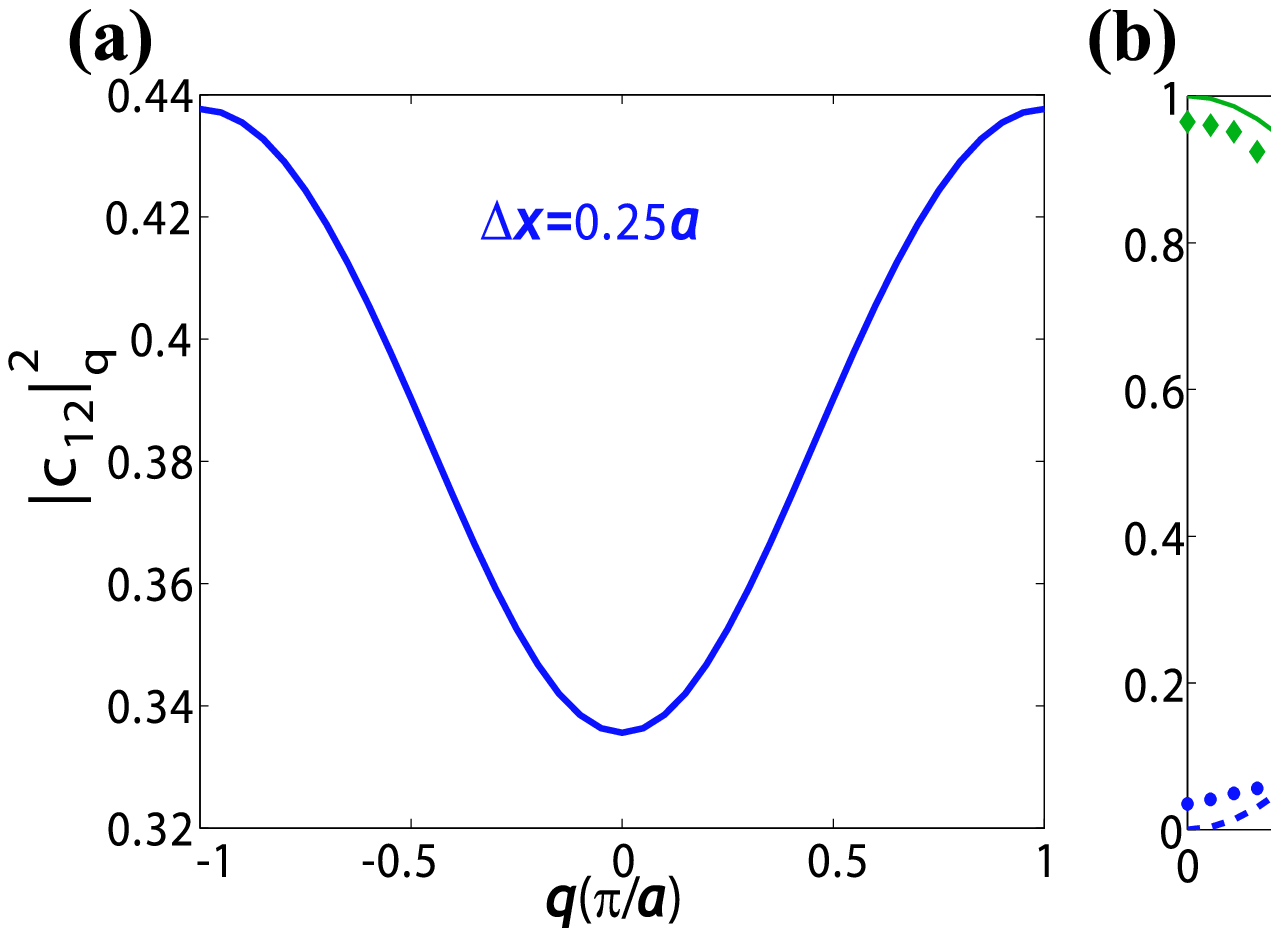}
\caption{(Color online) (a)$|c_{12}|_{q}^2=|\bra \Psi_{n=2,q}(x)|\Psi_{1,q}(x-\Delta x) \ket|^{2}$ as a function of quasi-momentum $q$ for a single-step pulse (single lattice displacement) with the optimum displacement of $\Delta x = 0.25a$ for the lattice depth of $U_{\circ}=18E_{R}$. (b) $P_{1n}=( 1/2\pi a) \int_{-\pi/a}^{\pi/a} d q |c_{1n}|_{q}^2$ as a function of displacement for bands $n=1$ and $n=2$ for the same lattice depth. For comparison, the measured values are shown on the same plot.}
\label{fig9}
\end{figure}

\begin{figure}[b]
			\includegraphics[scale=0.25]{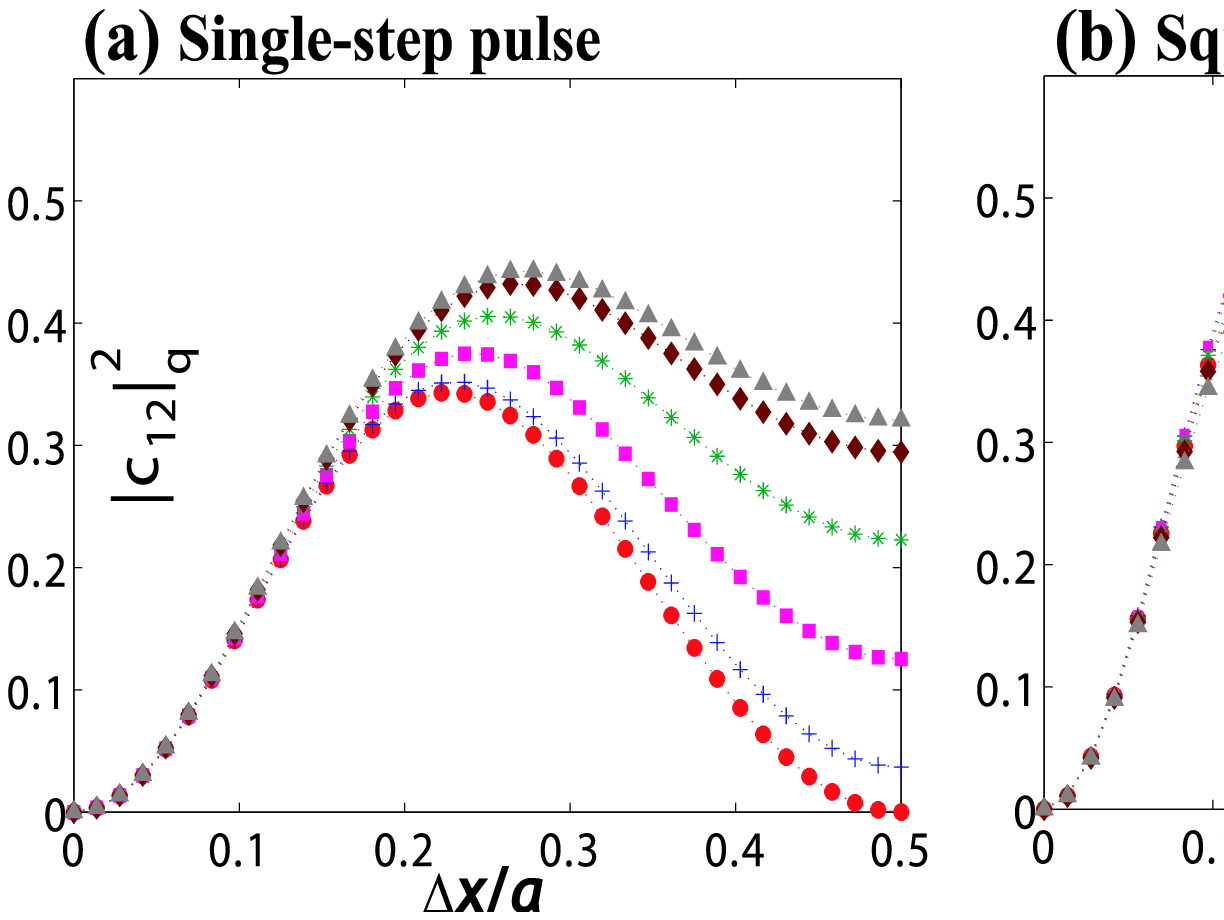}
\caption{(Color online) $|c_{12}|_{q}^2$, the coupling probability due to (a) single-step pulse, and (b) square pulse($\tau=0.35$) for different quasi-momentum components as a function of displacement.}
\label{fig10}
\end{figure}

Figure~\ref{fig9}(a) shows $|c_{12}|_{q}^2$, the coupling probability of the lowest two bands as a function of quasi-momentum for the lattice depth of $U_{\circ}=18E_{R}$. Figure~\ref{fig9}(b) is $P_{1n}$ , coupling probability of the lowest band $n=1$ with itself and with the second band $n=2$ as a function of displacement. At this lattice depth, the coupling probability of the lowest two bands for the optimum displacement of $\Delta x = 0.25a$, with $a$ the lattice spacing is $P_{12}=0.39$ compared to the maximum coupling of $1/e \approx~0.37$ for a harmonic potential. For comparison with the model, the measured values are shown on Fig.~\ref{fig9}(b). The deviation of measured value for $P_{11}$ from $1$ and the measured value for $P_{12}$ from $0$ at $\Delta x /a=0$ is due to transfer of atoms to the second band during the filtering stage (see section~\ref{experiment} and \cite{myrskog}), which is done for the purpose of measuring populations of the two lowest bands.

Each quasi-momentum contributes differently to the total coupling probability.   
Figure~\ref{fig10} shows the coupling probability $|c_{12}|_{q}^2$ for different single quasi-momentum components after single-step(a) and square(b) pulses as a function of displacement. 

\begin{figure}[t]
			\includegraphics[scale=0.25]{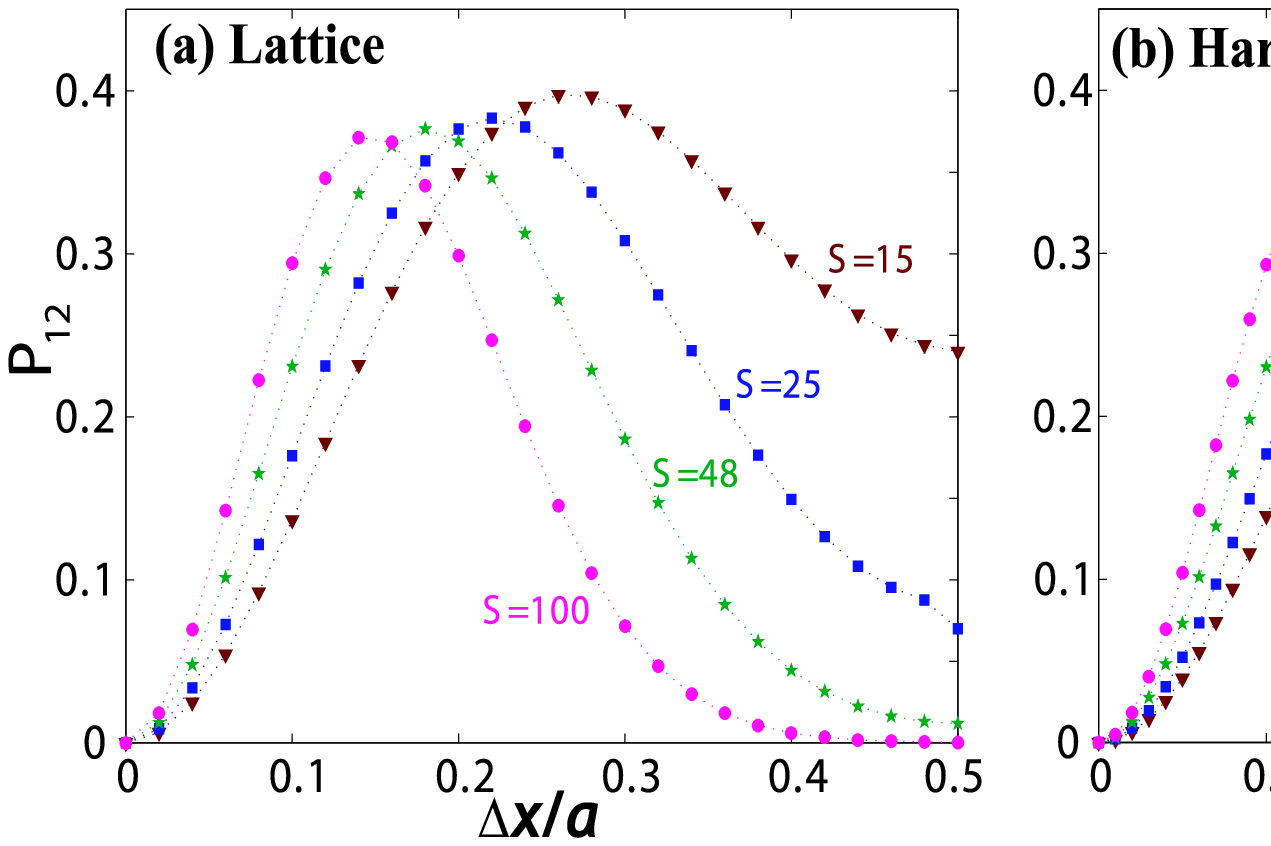}
\caption{(Color online) (a) Optimum $P_{12}$ after a single-step pulse as a function of displacement for different lattice depths $U_{\circ}=sE_{R}$. (b) The same overlap for harmonic potentials with $\omega_{ho}=\overline{\omega}_{12}$, where $\overline{\omega}_{12}$ is the average frequency difference of the two lowest bands in a lattice depth of $U_{\circ}=sE_{R}$. Displacement in the case of harmonic potential is expressed in units of lattice spacing $a = \pi s^{1/4} \sigma$, with $\sigma=\sqrt{\hbar/m\omega_{ho}}$.} 
\label{fig11}
\end{figure}

The coupling probability changes as the lattice depth is changed. Figure~\ref{fig11}(a) is the coupling probability $P_{12}$ due to a single-step pulse as a function of displacement for different lattice depths. As the lattice depth is increased, the maximum coupling decreases and approaches the harmonic limit of $1/e$ at large lattice depths. 
To see the effect of anharmonicity of the lattice potential, we calculate the coupling probability for the harmonic potential.

Figure~\ref{fig11}(b) is a plot of coupling probability $P_{12}$ as a function of displacement for different harmonic potentials. The frequency of the harmonic potentials have been chosen to match the average frequency of the lowest two bands of a particular lattice depth, i.e., we set $\omega_{ho}=\overline{\omega}_{12}$ of the lattice. The maximum coupling in the harmonic case is $1/e$ and it is the same for different harmonic frequencies, but the magnitude of optimum displacement decreases as the harmonic frequency is increased. The coupling probability between the ground and excited states in a harmonic potential after displacement of the ground state can be calculated analytically. Using the momentum representation of wavefunctions for the ground and the first excited states, the coupling after a single displacement of the ground state wavefunction is $\bra 1|\widehat{D}(q)|0\ket = \int^{+\infty}_{-\infty}\phi^{*}_{1}(p)e^{-\frac{ipq}{\hbar}}\phi_{0}(p)dp = \xi e^{-\frac{1}{2}\xi^{2}}$. Here, $\widehat{D}(q)$ is the harmonic oscillator displacement operator,
$\xi=\frac{p_{0}q}{\hbar}$ is a dimensionless displacement parameter;  $p_{0}=\sqrt{\frac{\hbar m\omega}{2}}$ is the width of the ground state wavefunction in momentum space, $\phi_{0}(p)=\frac{1}{(2\pi)^{1/4}p^{1/2}_{0}}e^{-(\frac{p}{2p_{0}})^{2}}$, and  
$\phi_{1}(p)=\frac{i}{(2\pi)^{1/4}p^{1/2}_{0}}(\frac{p}{p_{0}})e^{-(\frac{p}{2p_{0}})^{2}}$ are the momentum-space wavefunctions of the ground and first excited states of the harmonic oscillator.    
Then, the coupling probability is $|\langle 1|\widehat{D}(q)|0\rangle|^{2}=\xi^{2}e^{-\xi^{2}}$, which has a maximum of $1/e$ when $|q|=\frac{\hbar}{p_{0}}=2x_{0}$, where $x_{\circ}=\sqrt{\hbar/2m\omega}$ is the width of the ground state wavefunction in real space. In fact, in the harmonic oscillator case any combination of displacements and time evolutions results in the maximum coupling of $1/e$. Since the optimum displacement $x_{0}$ is proportional to $\omega^{-1/2}$ as the harmonic frequency is increased the optimum displacement decreases. This general trend remains valid in the lattice. In the harmonic approximation of the periodic potential, the lattice potential $U \propto \omega^2$, then the optimum displacement is proportional to $U^{-1/4}$. Therefore, as the potential depth is increased the optimum displacement decreases accordingly.

\begin{figure}[t]
			\includegraphics[scale=0.25]{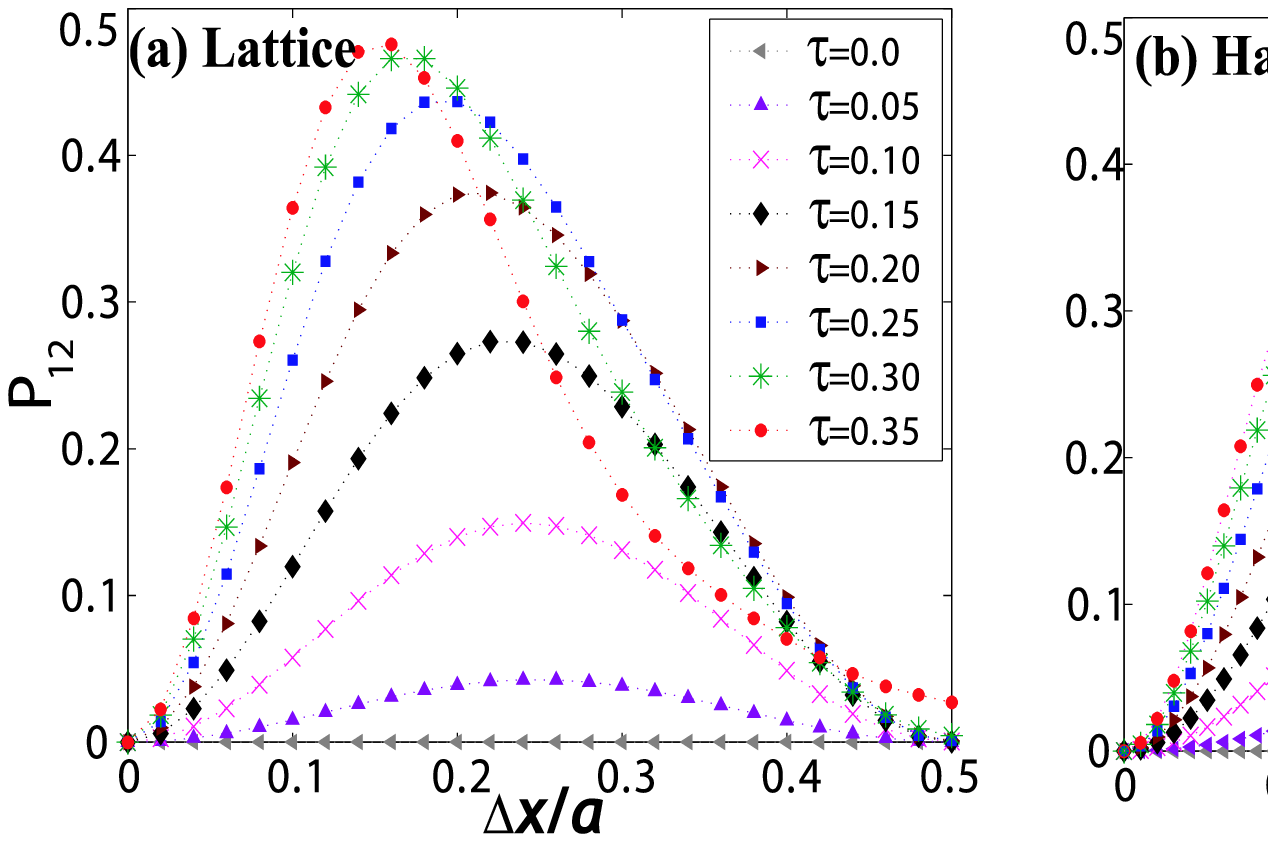}
\caption{(Color online) (a) $P_{12}$ after a square pulse as a function of displacement for different delays in a lattice of depth $U_{\circ}=18E_{R}$. $\tau= {\rm delay} /\overline{T}_{12}$ is indicated in the legend for each curve. The optimum scaled delay between the two displacements for this lattice depth is $\tau=0.35$ ( for $U_{\circ}=18E_{R}$, $\overline{T}_{12}=200 \mu s$, and optimum delay $=70\mu s$ ). 
(b) $P_{12}$ for the harmonic oscillator potential with $\omega_{ho}= \overline{\omega}_{12}$ of the lattice. The delays in (b) are the same as in (a). Displacement in the case of harmonic potential is expressed in terms of lattice spacing $a = \pi s^{1/4} \sigma$, with $\sigma=\sqrt{\hbar/m\omega_{ho}}$.}
\label{fig12}
\end{figure}

To find the optimum parameters of a square pulse, we calculate the coupling probability with respect to the two parameters of the pulse, i.e., magnitude of the displacement and the time delay between the two displacements. The maximum overlap for a specific displacement is found for different delays and the same calculation is repeated for different displacements to find the optimum displacement and the optimum delay. Figure~\ref{fig12}(a) is a plot of $P_{12}$, the coupling probability of the lowest two bands as a function of displacement for different delays. For a lattice depth of $U_{\circ}=18E_{R}$ we find the maximum coupling probability of $P_{12}=0.477$ for the optimum displacement of $\Delta x = 0.154a$ and the optimum delay of $0.35 \overline{T}_{12}$, a $22\%$ fractional improvement over the optimized single-step pulse (Table~\ref{tb1}). Again, we compare the lattice case with the harmonic potential. Figure~\ref{fig12}(b) is the plot of $P_{12}$ due to square pulse as a function of displacement for a harmonic potential. In the case of a harmonic potential, the maximum coupling of $1/e$ stays the same for different optimum displacements and delays. 
Comparisons to harmonic potential in both cases of single-step and square pulses suggest that the anharmonicity is the main reason for higher couplings in lower lattice depths. 

The interesting parameter in the square pulse is the delay, as this signifies how the phase difference between different states affects the coupling between the lowest two bands. The amount of phase accumulated during the time evolution in the lattice depends on the energy, and therefore the lattice depth. We calculate how the optimum delay changes as the lattice depth is changed. As the natural time scale of the system we choose $\overline{T}_{12}$, the average period of oscillations between the lowest two bands and present the scaled delay $\tau$ as the ratio delay$/\overline{T}_{12}$.

\begin{figure}[t]
\includegraphics[scale=0.30]{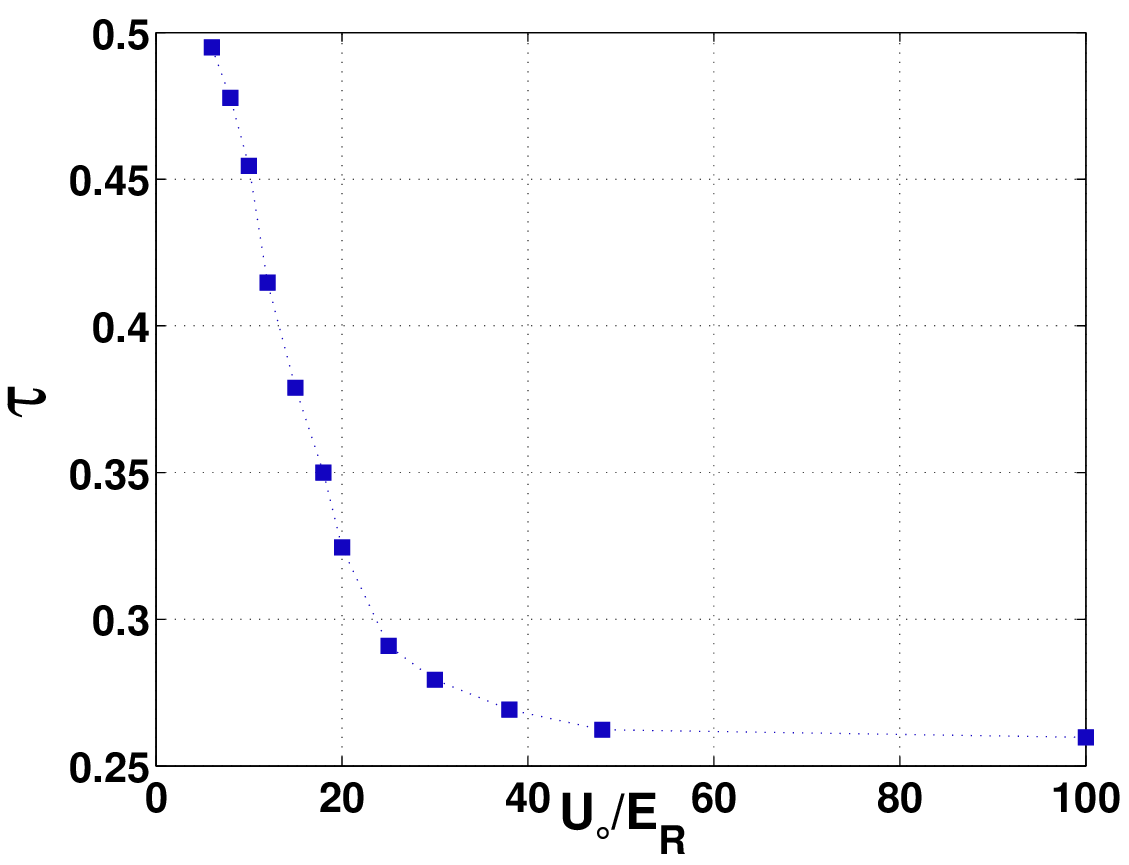}
\caption{ (Color online) The optimum smallest scaled delay of the square pulse, $\tau = {\rm delay} /\overline{T}_{12}$ as a function of lattice depth. The optimum displacement is different for different lattice depths.}
\label{fig13}
\end{figure} 
  
\begin{figure}[b]
\includegraphics[scale=0.25]{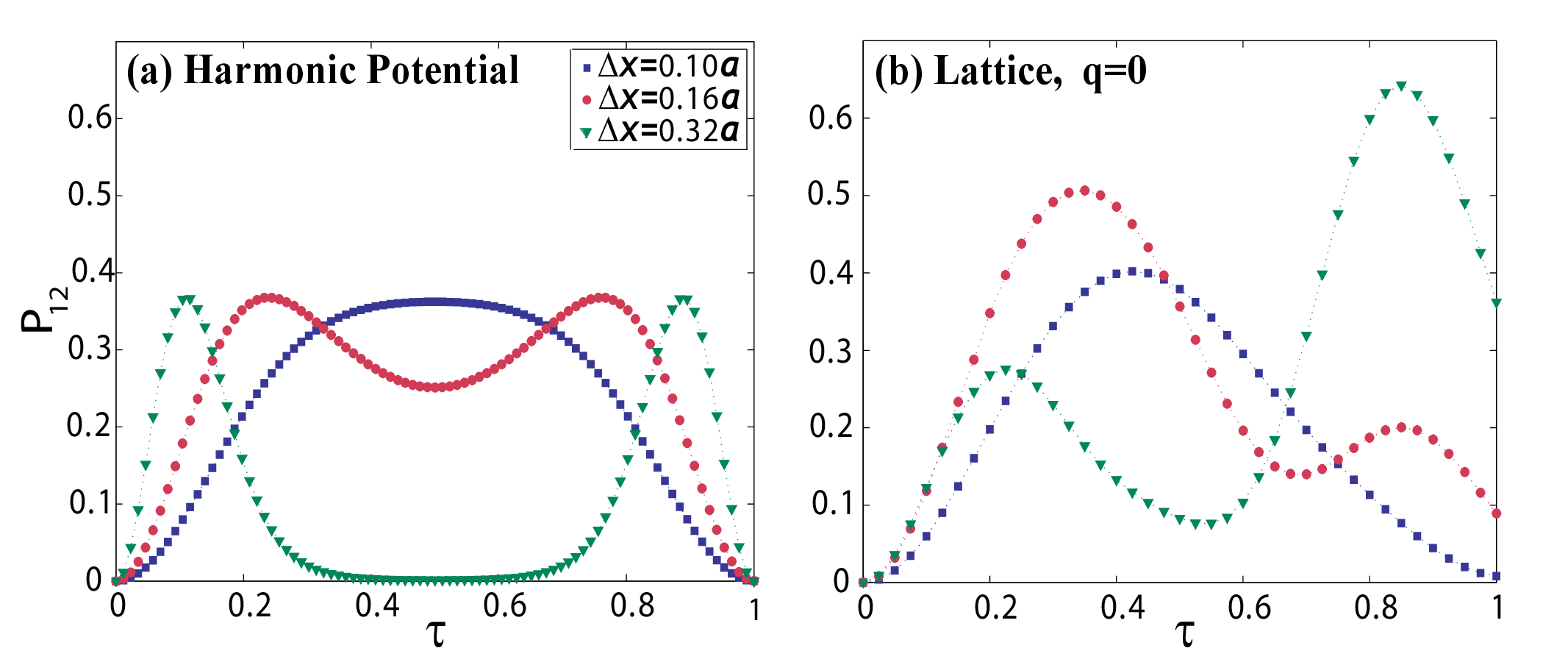}
\caption{(Color online) (a) $P_{12}$ after a square pulse as a function of scaled delay $\tau$ for a harmonic potential with $\omega_{ho}= \overline{\omega}_{12}$ in a lattice depth of $U_{\circ}=18E_{R}$. Depending on the magnitude of displacement, the optimum delay changes but the maximum overlap $P_{12}$ stays the same.
(b) $P_{12}$ after a square pulse as a function of scaled delay $\tau$ for a single quasi-momentum component $q=0$ of the same lattice depth. $P_{12}$, the maximum overlap is different for different displacements. Displacements in (b) are the same as in (a).}
\label{fig14}
\end{figure}

Figure~\ref{fig13} is a plot of the smallest optimum scaled delay $\tau$ as a function of lattice depth. As the lattice depth is increased the scaled delay approaches $0.25  \overline{T}_{12}$.
To have a better understanding of the optimum delay for deeper lattices, we calculate the optimum parameters of a square pulse for a harmonic potential with frequency $\omega_{ho}= \overline{\omega}_{12}$ of the lattice. Figure~\ref{fig14}(a) is a plot of the coupling $P_{12}$ as a function of scaled delay $\tau$ for three different fixed displacements of the harmonic potential. In the case of harmonic potential the maximum coupling stays the same for different displacements at optimum delays. In addition there is no gain in coupling due to the square pulse over the single-step pulse. To see the effects of anharmonicity alone in the case of lattice potential, we consider only one quasi-momentum component and observe how the coupling between the lowest two states changes as a function of delay. The plot in Fig.~\ref{fig14}(b) shows $|c_{12}|_{q=0}^2$ as a function of $\tau$ for the lattice depth of $U_{\circ}=18E_{R}$. Similar to the harmonic potential, the optimum delay is different for different displacements, but unlike the harmonic case, the maximum coupling changes for different displacements. 

\begin{figure}[t]
\includegraphics[scale=0.30]{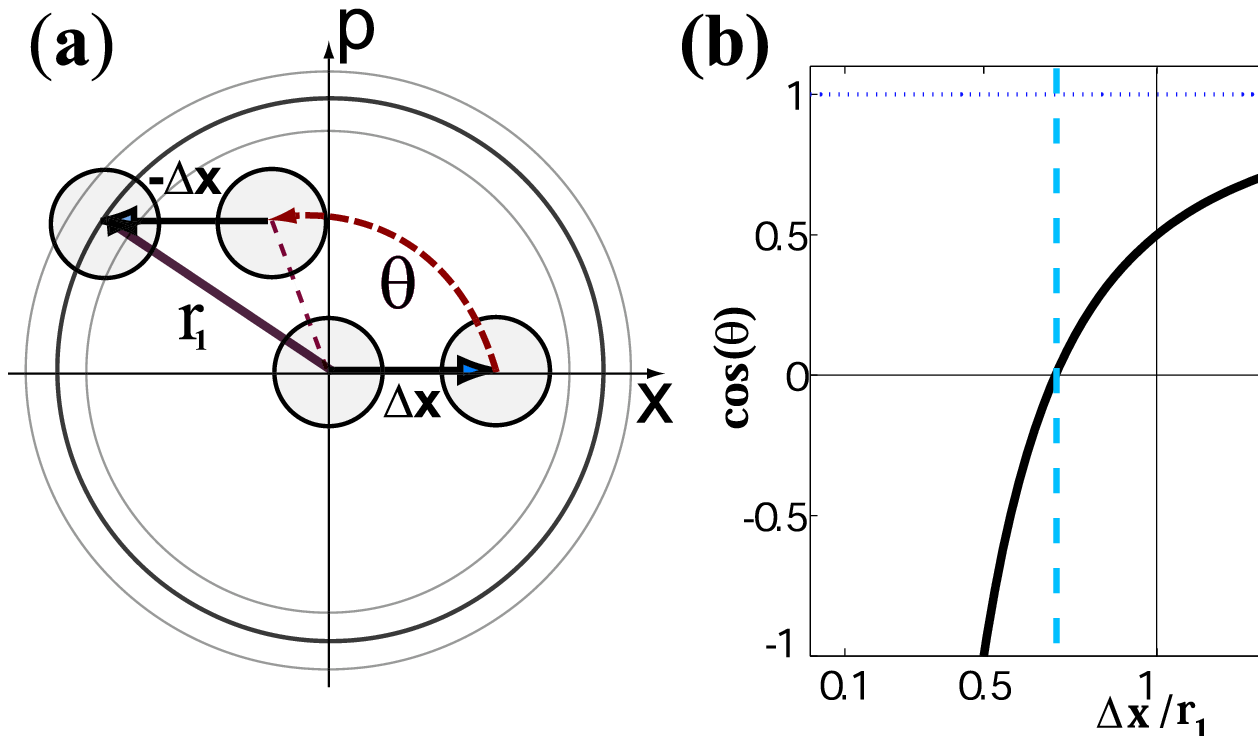}
\caption{ (Color online) (a) Phase-space schematics (not to scale) for the harmonic oscillator illustrating the square pulse sequence. The optimum delay $t$ ($\theta = \omega_{ho} t$) depends on the magnitude of displacement. (b) $cos(\theta) = 1-1/(2 (\Delta x /r_{1})^2$ as a function of $(\Delta x/r_{1})$; $r_{1}$ is the phase-space radius of the first excited state of the harmonic oscillator. For the displacement of $\Delta x =0.5r_{1}$ the optimum delay is $t=T/2$ for $cos(\theta)=-1$. As the displacement is increased the optimum delay becomes smaller; at $\Delta x=r_{1}/\sqrt{2}$ (the blue dashed line), the delay of $t=T/4$ for $\theta=\pi/2$ results in maximum coupling.}
\label{fig15}
\end{figure}

For a harmonic oscillator, one would intuitively expect the optimum delay to be $T/2$, half the period, since a displacement plus a delay of $T/2$ and a second displacement in the opposite direction is equivalent to a large single displacement. We draw a schematic of the phase-space representation of the harmonic oscillator number states in Fig.~\ref{fig15}(a) to illustrate the effect of the displacement and time-evolution operators in coupling the ground state to the first excited state. In this representation, each number state is represented as a circle of radius $r_{n}=\sqrt{n+1/2}$ with the approximate width of $ 1/2 \sqrt{n}$. Since the number states are complete the circles fill the phase-space. For sake of illustration the radii and the widths in Fig.~\ref{fig15}(a) are not drawn to scale. In this figure, the ground state corresponds to the central circle and the circular band represents the first excited state. A displacement corresponds to a translation along the x-axis and a delay or time evolution is a rotation in phase-space by an angle $\theta =\omega t$.      
As figure~\ref{fig15}(a) shows, the optimum delay depends on the magnitude of displacement. In fact, for a displacement larger than $r_{1}/2$ the optimum delay is smaller than $T/2$.
Figure~\ref{fig15}(b) is a plot of $cos(\theta)=1-1/(\Delta x /r_{1})^2$ (calculated geometrically) as a function of displacement. The delay is $t=T/2$ only for a displacement of $\Delta x =0.5r_{1}$. As the displacement is increased from $r_{1}/2$ the optimum delay becomes smaller; for example at $\Delta x=r_{1}/\sqrt{2}$ the delay of $t=T/4$ for $cos(\theta)=0$ results in maximum coupling. 

\begin{figure}[t]
\mbox{
\includegraphics[scale=0.35]{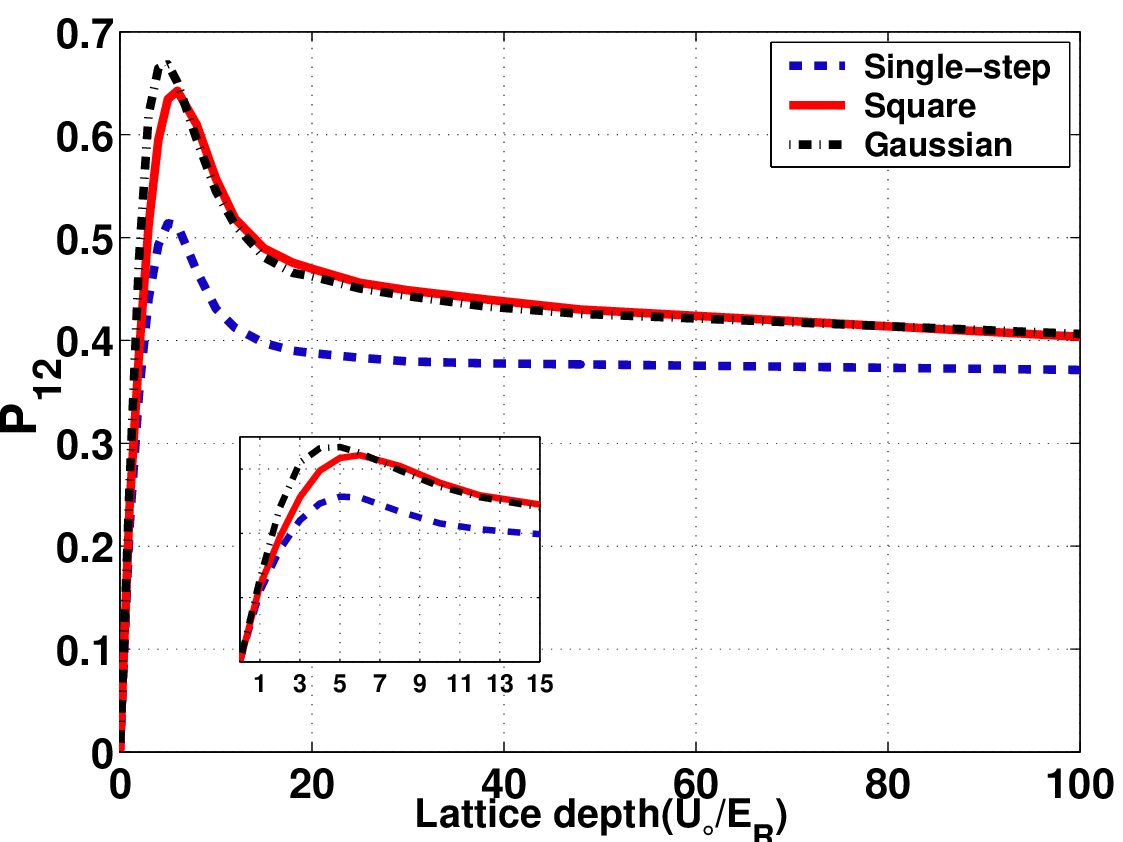}
}
\caption{(Color online) Calculated $P_{12}$ as a function of lattice depth for optimized single-step, square and Gaussian pulses. The coupling probability increases as the lattice depth is decreased until the second band is no longer supported, at which part it drops off sharply. The inset shows $P_{12}$ for lattice depths less than $15E_{R}$. }
\label{fig16}
\end{figure}

Finally we compare the coupling probability for different lattice depths due to optimized single-step, square and Gaussian pulses. We find that the square pulse yields a larger coupling probability than the single-step pulse for all lattice depths, with a fractional improvement as large as $30\%$ for a lattice depth of $U_{\circ}=8E_{R}$. The coupling probability between the lowest two bands, $P_{12}$, is shown as a function of lattice depth in Fig.~\ref{fig16} for each of the three pulses.
The effect of the Gaussian pulse is very similar to that of the square pulse except for lattice depths lower than $7E_{R}$, where the Gaussian pulse results in slightly larger coupling probability. For all pulses, the coupling increases as the lattice depth is decreased, until the second band becomes completely unbound, at which point it drops off sharply. The coupling probability for a single-step pulse increases from the harmonic potential limit of $1/e$ for large lattice depths to the maximum coupling of $0.51$ for $U_{\circ}=5E_{R}$. The square and Gaussian pulses perform even better, yielding couplings of $0.64$ at $6E_{R}$ and $0.67$ at $5E_{R}$, respectively.

\section{Conclusion}
We have compared the effectiveness of three different pulses at coupling atoms in the lowest two bands of a far-detuned 1D optical lattice.  The three pulses are time-dependent lattice displacements following single-step, square and Gaussian profiles in time. We find experimentally that the square pulse results in larger coupling probability than the single-step and Gaussian pulses, a result which is in agreement with simulations. The maximum coupling probability observed experimentally, for a square pulse in a lattice of depth $U_{\circ}=18E_{R}$, is $0.33 \pm 0.01$; this is far higher than in any previous experiments, but lower than the theoretical maximum coupling of $0.477$. We believe that this is due to lattice inhomogeneity. 
We have also investigated optimizing pulses for inverting the populations of the lowest two bands in a pulse-echo experiment. We find that the square and Gaussian pulses result in larger echo amplitudes -- $0.23$ and $0.24$ times the amplitude of initial oscillations, respectively -- than that arising from a single-step pulse, which is $0.19$. The advantage of the Gaussian pulse arises because it leads to less loss from the lattice. For the optimized echo square pulse, when starting with atoms in the lowest band $78\%$ of the remaining atoms are predicted to be transferred to the second band. Experimentally, $63\%$ is our maximum observed value.   
We find that the agreement between experimental and calculated optimized pulse parameters in the echo experiment is not as good as the agreement in the case of the coupling experiment. The main difference between the two problems is that the echo pulse needs to efficiently invert populations from an arbitrary initial superposition of the two states. In addition, the pulses in the echo experiment have been optimized with respect to one parameter only, the echo amplitude for a particular initial coherence.  A complete optimization of the pulses will require full quantum process tomography \cite{myrskog}. 

We have experimentally observed normalized echo amplitudes of up to $0.24$ in a lattice of depth $20E_{R}$, more than an order of magnitude larger than the echo amplitudes observed previously for the motional states of atoms in optical lattices \cite{Birkl}. Our numerical calculations show the coupling between the lowest two bands is more efficient in shallow lattices and should reach a maximum value of $0.67$ for a Gaussian pulse when the lattice depth is $5E_{R}$, while a square pulse results in a maximum coupling of $0.64$ for a lattice depth of $6E_{R}$. The square pulse outperforms the single-step pulse for all lattice depths and it results in larger couplings than the Gaussian pulse for lattice depths larger than $7 E_{R}$.  This is an example of coherent control, the optimized time delay introducing a relative phase which maintains constructive interference into the desired final state and destructive interference into the lossy states. These improved pulses have enabled us to extend pulse echo work to longer times, discovering a coherence plateau out to $25$ oscillation periods, a topic of ongoing study.

Future work will include combining amplitude and phase modulation of the lattice potential to achieve better coupling and lower loss, investigation of adiabatic passage approaches, as well as the application of optimal control theory to find improved pulse shapes and parameters.  The ability to efficiently couple quantized states in an optical lattice is crucial for state preparation as well as for the use of pulse echoes to preserve intra-well coherence, and similar pulses should find application in other systems described by tilted-washboard potentials.

\section{Acknowledgments}

We would like to thank Daniel James and Ardavan Darabi for useful discussions and we acknowledge financial support from NSERC, CIAR, PRO, and the DARPA QuIST program (managed by the AFOSR under agreement no. F49620-01-1-0468).

\end{document}